\documentclass[12pt]{article}
\usepackage{amssymb,amsmath}
\usepackage{comment}
\usepackage{graphicx}
\usepackage{bm}
\usepackage{enumerate}
\usepackage{subfigure}
\usepackage{here}
\usepackage{hyperref}

\numberwithin{equation}{section}

\begin{document}

\thispagestyle{empty}
\begin{flushright}
CALT-TH-2014-159\\
OU-HET 833\\

\end{flushright}
\vskip1cm
\begin{center}
{\Large  Membrane Quantum Mechanics}

\vskip1.5cm
Tadashi Okazaki\footnote{tadashi@theory.caltech.edu} 

\bigskip\bigskip
{\it 
Walter Burke Institute for Theoretical Physics\\
California Institute of Technology \\
Pasadena, CA 91125, USA}
\\
\vskip1mm
{\it and}
\\
\vskip3mm
{\it Department of Physics, Graduate School of Science \\
Osaka University \\
Toyonaka, Osaka 560-0043, Japan}

\end{center}

\vskip1.5cm
\begin{abstract}
We consider the multiple M2-branes wrapped on a 
compact Riemann surface and study the arising quantum mechanics 
by taking the limit where the size of the Riemann surface goes to zero. 
The IR quantum mechanical models resulting 
from the BLG-model and the ABJM-model compactified on 
a torus are $\mathcal{N}=16$ and $\mathcal{N}=12$ 
superconformal gauged quantum mechanics. 
After integrating out the auxiliary gauge fields 
we find $OSp(16|2)$ and $SU(1,1|6)$ quantum mechanics 
from the reduced systems.  
The curved Riemann surface is taken as a 
holomorphic curve in a Calabi-Yau space 
to preserve supersymmetry 
and we present a prescription of the topological twisting. 
We find the $\mathcal{N}=8$ superconformal 
gauged quantum mechanics that may describe the motion 
of two wrapped M2-branes in a K3 surface. 
\end{abstract}


\newpage
\setcounter{tocdepth}{2}
\tableofcontents

\section{Introduction}
M2-brane appears to be a fundamental object in M-theory 
in the sense that it can be identified with the fundamental string 
after the compactification of M-theory 
to type IIA string theory \cite{Duff:1987bx}. 
In the past decade some progress has been made 
in finding the low-energy world-volume descriptions for 
multiple M2-branes. 
Inspired by the work in \cite{Schwarz:2004yj} 
and \cite{Basu:2004ed}, 
Bagger, Lambert and Gustavsson discovered 
the three-dimensional $\mathcal{N}=8$ 
superconformal Chern-Simons-matter theory, 
the so-called BLG-model
\cite{Bagger:2006sk,Bagger:2007jr,Bagger:2007vi,
Gustavsson:2007vu,Gustavsson:2008dy}. 
Subsequently Aharony, Bergman, Jafferis and Maldacena constructed 
the three-dimensional $\mathcal{N}=6$ 
superconformal Chern-Simons-matter theory, 
the so-called ABJM-model \cite{Aharony:2008ug}. 
Since then the BLG-model and the ABJM-model 
have been proposed as the low-energy effective world-volume theories of 
multiple planar M2-branes.

In this paper we study more general M2-branes 
wrapping a compact Riemann surface $\Sigma_{g}$ of genus $g$. 
For $g\neq 1$ the world-volume of the M2-branes is curved 
and the Riemann surface has to be taken 
as a holomorphic curve in a Calabi-Yau manifold 
to preserve supersymmetry. 
The construction of such world-volume theories on the wrapped branes 
can be implemented 
as topologically twisted theories \cite{Bershadsky:1995qy}. 
For the world-volume descriptions of wrapped M2-branes, 
we can take the further limit 
where the energy scale is much smaller than 
the inverse size of the Riemann surface. 
This implies that the Riemann surface shrinks to zero and thus 
the three-dimensional world-volume theories reduce 
to a one-dimensional field theories, i.e. quantum mechanics. 
The purpose of the present paper is to derive and study 
the emerging IR quantum mechanics by reducing 
the BLG-model and the ABJM-model. 

It has been argued in \cite{Gauntlett:2001qs} that 
there exist IR fixed points with AdS$_{2}$
factors in $d=11$ supergravity solutions 
describing the M2-branes wrapping $\Sigma_{g}$ 
which are gravity dual 
to superconformal quantum mechanics (SCQM). 
Quite interestingly we show that our low-energy effective quantum mechanics 
possesses a one-dimensional superconformal symmetry. 
Generally superconformal quantum mechanics is characterized 
by a supergroup that contains 
a one-dimensional conformal group $SL(2,\mathbb{R})$ 
and an R-symmetry group as factored bosonic subgroups. 
The first detailed analysis for a simple conformal quantum mechanical
model, the so-called DFF-model is found  
in \cite{deAlfaro:1976je} 
and there has been a number of attempts to 
construct superconformal mechanics 
since the earliest work of \cite{Akulov:1984uh,Fubini:1984hf}. 
One of the most powerful way to build such 
superconformal quantum mechanics is to resort to 
superspace and superfield formalism.  
However, it is 
unreasonable and unsuccessful for highly supersymmetric cases 
with $\mathcal{N}>8$ supersymmetry 
because it is extremely difficult to pick up 
irreducible supermultiplets by imposing the appropriate 
constraints on the superfields 
\cite{Bellucci:2004ur}. 
Remarkably we find that 
such highly extended superconformal quantum mechanical models 
arise from the M2-branes wrapping a torus 
and that our reduced quantum mechanical actions 
agree with the predicted form for $\mathcal{N}>4$ SCQM in 
\cite{Ivanov:1988it,Wyllard:1999tm,Fedoruk:2011aa}.

This paper is organized as follows. 
In section \ref{secm2} we review the BLG-model 
\cite{Bagger:2006sk,Bagger:2007jr,Bagger:2007vi,
Gustavsson:2007vu,Gustavsson:2008dy} 
and the ABJM model \cite{Aharony:2008ug}. 
In section \ref{secflat} we study the 
multiple M2-branes wrapped around a torus. 
From the BLG-model we find that the 
low-energy dynamics is described by $\mathcal{N}=16$ superconformal 
gauged quantum mechanics. 
Furthermore we show that 
$OSp(16|2)$ superconformal quantum mechanics appears from 
the reduced system after integrating out the auxiliary gauge field. 
Similarly from the ABJM-model 
$\mathcal{N}=12$ superconformal gauged quantum mechanics  
makes an entrance at low-energy 
and we find the reduced quantum mechanics with $SU(1,1|6)$ symmetry. 
In section \ref{seccurv} 
we clarify the description for curved M2-branes 
wrapping a holomorphic curve in a Calabi-Yau manifold. 
We discuss the amount of preserved supersymmetries 
and establish a prescription for the topological twisting. 
In section \ref{cy2sec} 
we examine the two M2-branes wrapped on a 
Riemann surface of genus $g>1$ embedded in a K3 surface in detail. 
%
Finally in section \ref{secconc} 
we conclude and discuss some directions for future research.

\section{World-volume theories of M2-branes}\label{secm2}
\subsection{BLG-model}
\label{secm2a}
The BLG-model is a three-dimensional 
$\mathcal{N}=8$ superconformal 
Chern-Simons-matter theory proposed as 
a low energy world-volume theory of multiple M2-branes
\cite{Bagger:2006sk,Bagger:2007jr,Bagger:2007vi,
Gustavsson:2007vu,Gustavsson:2008dy}. 
It is based on a 3-algebra $\mathcal{A}$, 
which is an $N$ dimensional vector space 
endowed with a trilinear skew-symmetric product $[A,B,C]$ 
satisfying 
\begin{align}
\label{fundid1}
[A,B,[C,D,E]]
=[[A,B,C],D,E]+[C,[A,B,D],E]+[C,D,[A,B,E]].
\end{align}
This is called the fundamental identity and  
extends the Jacobi identity 
of Lie algebras to the 3-algebras. 
If we let $T^{a}$, $a=1,\cdots,N$ 
be a basis of the algebra, 
the 3-algebra is specified 
by the structure constants ${f^{abc}}_{d}$ 
\begin{align}
\label{3algstc1}
[T^{a},T^{b},T^{c}]={f^{abc}}_{d}T^{d}.
\end{align}
With the structure constant, 
the fundamental identity (\ref{fundid1}) can be expressed as
\begin{align}
\label{gomefundid2}
{f^{abg}}_{h}{f^{cde}}_{g}
={f^{abc}}_{g}{f^{gde}}_{h}
+{f^{abd}}_{g}{f^{cge}}_{h}
+{f^{abe}}_{g}{f^{cdg}}_{h}
\end{align}
Classification of the 3-algebras $\mathcal{A}$ 
requires finding the solutions 
to the fundamental identity (\ref{gomefundid2})
for the structure constants ${f^{abc}}_{d}$. 

In order to derive the equations of motion 
of the BLG-model from a Lagrangian description, 
a bi-invariant non-degenerate metric $h^{ab}$ 
on the 3-algebra $\mathcal{A}$ is needed. 
Bi-invariance requires the metric to satisfy 
${f^{abc}}_{e}h^{ed}+{f^{bcd}}_{e}h^{ae}=0$. 
This implies that 
the tensor $f^{abcd}\equiv {f^{abc}}_{e}h^{ed}$ is totally
anti-symmetric. 
The metric $h^{ab}$ arises by postulating a non-degenerate, 
bilinear scalar product $\mathrm{Tr} (\ ,\ ) $ 
on the algebra $\mathcal{A}$:
\begin{align}
\label{3algmet}
h^{ab}=\mathrm{Tr}(T^{a},T^{b}).
\end{align}
The Lagrangian of the BLG-model is specified by the 
structure constant ${f^{abc}}_{d}$ 
and the bi-invariant metric $h^{ab}$.

The field content of the BLG-model is 
eight real scalar fields $X^{I}=X_{a}^{I}T^{a}, I=1,\cdots,8$ 
, fermionic fields 
$\Psi_{\dot{A}a}=\Psi_{\dot{A}a}T^{a}, \dot{A}=1,\cdots,8$ 
and non-propagating gauge fields $A_{\mu ab},\mu=0,1,2$. 
The bosonic scalar fields $X^{I}$ and the fermionic fields
$\Psi_{\dot{A}}$ are 
$\bm{8}_{v}$ and 
$\bm{8}_{c}$ 
of an $SO(8)$ R-symmetry respectively. 
Also they are the fundamental representations of the 3-algebra. 
Gauge fields $A_{\mu ab}$ are the 3-algebra valued
world-volume vector fields. 
They are anti-symmetric 
under two indices $a,b$ of the 3-algebra 
$A_{\mu ab}=-A_{\mu ba}$. 
 
$\Psi_{\dot{A}a}$ is defined as an $SO(1,10)$ Majorana fermion and its
conjugate is given by
\begin{equation}
 \overline{\Psi}:=\Psi^{T}\mathcal{C}
\end{equation}
where $\mathcal{C}$ is the $SO(1,10)$ charge conjugation matrix satisfying
\begin{equation}
 \mathcal{C}^{T}=-\mathcal{C},\ \ \ \mathcal{C}\Gamma^{M}\mathcal{C}^{-1}=-(\Gamma^{M})^{T}.
\end{equation}
Gamma matrix $\Gamma^{M}$ is the representation of 
the $SO(1,10)$ Clifford algebra
\begin{equation}
\label{10gamma}
 \{\Gamma^{M},\Gamma^{N}\}=2\eta^{MN},\ \ \  \Gamma^{10}:=\Gamma^{0\cdots 9}
\end{equation}
where $\eta^{MN}=\textrm{diag}(-1,+1,+1,\cdots,+1)$. 
$\Gamma^{M}$ can be decomposed as
\begin{align}
\label{11gamma}
\begin{cases}
 \Gamma^{\mu}=\gamma^{\mu}\otimes\tilde{\Gamma}^{9}&\mu=0,1,2\cr
 \Gamma^{I}=\mathbb{I}_{2}\otimes\tilde{\Gamma}^{I-2}&I=3,\cdots,10\cr
\end{cases}
\end{align}
where 
\begin{equation}
\label{3dgamma}
 \gamma^{0}=\left(
\begin{array}{cc}
0&1\\
-1&0\\
\end{array}
\right)=i\sigma_{2},\ \ \ \gamma^{1}=\left(
\begin{array}{cc}
0&1\\
1&0\\
\end{array}
\right)=\sigma_{1},\ \ \ 
\gamma^{2}=\left(
\begin{array}{cc}
1&0\\
0&-1\\
\end{array}
\right)=\sigma_{3}
\end{equation}
and $\tilde{\Gamma}^{I}$ is the $SO(8)$ $16\times 16$ gamma matrix whose
chirality matrix is defined as 
$\tilde{\Gamma}^{9}:=\tilde{\Gamma}^{1\cdots 8}$. 
Correspondingly the charge conjugation matrix can be expanded as
\begin{align}
\mathcal{C}=\gamma^{0}\otimes \tilde{C}
\end{align}
where $\tilde{C}$ denotes 
the $SO(8)$ charge conjugation matrix satisfying 
\begin{align}
\tilde{C}^{T}=\tilde{C},\ \ \ \ \ 
\tilde{C}\tilde{\Gamma}^{I}\tilde{C}^{-1}
=-(\tilde{\Gamma}^{I})^{T}.
\end{align}
The fermionic field $\Psi$ is the real $\frac12 \cdot
2^{[\frac{11}{2}]}=32$-component Majorana spinor of eleven-dimensional
space-time obeying the chirality condition
\begin{equation}
\label{blgchiral}
 \Gamma^{012}\Psi=-\Psi.
\end{equation}
Although at this stage $\Psi$ has sixteen independent real components, 
they are reduced to eight when we treat it on-shell.
From (\ref{11gamma}) it follows that
\begin{equation}
 \Gamma^{012}=\Gamma^{34\cdots 10}=\mathbb{I}_{2}\otimes \tilde{\Gamma}^{9}
\end{equation}
and 
\begin{equation}
 \Gamma^{34\cdots 10}\Psi=-\Psi.
\end{equation}
This implies that 
$\Psi$ is the conjugate spinor representation $\bm{8}_{c}$ of the 
$SO(8)_{R}$ R-symmetry group.

The Lagrangian of the BLG-model is
\begin{align}
\label{blglagrangian}
 \mathcal{L}_{\textrm{BLG}}=&-\frac12 D^{\mu}X^{Ia}D_{\mu}X_{a}^{I}
+\frac{i}{2}\overline{\Psi}_{\dot{A}}^{a}\Gamma^{\mu}_{\dot{A}\dot{B}}D_{\mu}\Psi_{\dot{B}a}\nonumber\\
&+\frac{i}{4}\overline{\Psi}_{\dot{A}b}
\Gamma^{IJ}_{\dot{A}\dot{B}}X_{c}^{I}X_{d}^{J}\Psi_{\dot{B}a}f^{abcd}
-V(X)+\mathcal{L}_{\textrm{TCS}}
\end{align}
where 
\begin{align}
\label{blgpot}
 V(X)=&\frac{1}{12}f^{abcd}{f^{efg}}_{d}X_{a}^{I}X_{b}^{J}X_{c}^{K}
X_{e}^{I}X_{f}^{J}X_{g}^{K}\\
\label{blgcs}
 \mathcal{L}_{\textrm{TCS}}
=&\frac12 \epsilon^{\mu\nu\lambda}
\left(
f^{abcd}A_{\mu ab}\partial_{\nu}A_{\lambda cd}
+\frac23 {f^{cda}}_{g}f^{efgb}A_{\mu ab}A_{\nu cd}A_{\lambda ef}
\right).
\end{align}
The covariant derivative is defined as
\begin{equation}
\label{blgderiv}
 D_{\mu}X_{a}
:=\partial_{\mu}X_{a}-\tilde{A}_{\mu a}^{b}X_{b}
\end{equation}
where $\tilde{A}_{\mu b}^{a}:={f^{cda}}_{b}A_{\mu cd}$. 
Although the kinetic term of the gauge fields is similar to the conventional
Chern-Simons term, it is twisted by the structure constant of the 3-algebra. 
The gauge fields are non-propagating 
since they have at most first order derivative terms.

%
%
%
%
%
%
%
%
%
%
%
%
%
%
%
%
%
%
%
%
%
%
%

The supersymmetry transformations of the BLG-model are
\begin{align}
\label{blgsusy1}
\delta
 X_{a}^{I}&=i\overline{\epsilon}_{A}\Gamma^{I}_{A\dot{B}}\Psi_{\dot{B}a}\\ 
\label{blgsusy2}
\delta\Psi_{\dot{A}a}&=
D_{\mu}X_{a}^{I}\Gamma^{\mu}\Gamma^{I}_{\dot{A}B}\epsilon_{B}
-\frac16 X_{b}^{I}X_{c}^{J}X_{d}^{K}
{f^{bcd}}_{a}\Gamma^{IJK}_{\dot{A}B}\epsilon_{B}\\
\label{blgsusy3}
\delta\tilde{A}_{\mu a}^{b}
&=i\overline{\epsilon}_{A}\Gamma_{\mu}\Gamma^{I}_{A\dot{B}}
X_{c}^{I}\Psi_{\dot{B}d}{f^{cdb}}_{a}
\end{align}
where $\epsilon$ is the unbroken supersymmetry parameter obeying the chirality
condition
\begin{equation}
\label{blgchiralmtx}
 \Gamma^{012}\epsilon=\Gamma^{34\cdots 10}\epsilon=\epsilon.
\end{equation}
This means that $\epsilon$ transforms as the spinor representation
$\bm{2}$ of the $SL(2,\mathbb{R})$ 
and transforms as the spinor representation 
$\bm{8}_{s}$ of the $SO(8)_{R}$ R-symmetry. 
The action (\ref{blglagrangian}) is invariant 
under the supersymmetry transformations
(\ref{blgsusy1})-(\ref{blgsusy3}) 
up to a surface term.

If we assume that (i) the metric $h^{ab}$ of the 3-algebra $\mathcal{A}$ is 
 positive definite so that the kinetic term and the potential term are
 all positive, and that (ii) the dimension $N$ of the 3-algebra
 $\mathcal{A}$ is finite, then non-trivial 3-algebra $\mathcal{A}$ is
 uniquely determined as 
\cite{Papadopoulos:2008sk,Gauntlett:2008uf} 
\begin{align}
\label{a40a}
&f^{abcd}=\frac{2\pi}{k}\epsilon^{abcd}=:f\epsilon^{abcd}\\
\label{a40b}
&h^{ab}=\delta^{ab}
\end{align}
where the gauge indices 
$a,b,\cdots$ run from $1$ to $4$ 
and $k$ is the integer valued  Chern-Simons level.  
This is called the $\mathcal{A}_{4}$ algebra. 
For the $\mathcal{A}_{4}$ algebra 
one can realize two gauge groups, $SO(4)=SU(2)\times SU(2)/\mathbb{Z}_{2}$ 
and $Spin(4)=SU(2)\times SU(2)$ \cite{Lambert:2010ji}. 
The moduli space for $\mathcal{A}_{4}$ BLG-model with level $k$ 
is identified with \cite{Lambert:2010ji}
\begin{equation}
\label{blgmodulisp}
 \mathcal{M}_{k}=
\begin{cases}
\frac{\mathbb{R}^{8}\times \mathbb{R}^{8}}{D_{2k}}&\textrm{for}\ \ SO(4)\cr
\frac{\mathbb{R}^{8}\times \mathbb{R}^{8}}{D_{4k}}&\textrm{for}\ \ Spin(4).\cr
\end{cases}
\end{equation}
The limitation on the rank of the gauge algebra may only 
allow the BLG-model to describe two M2-branes 
in analogy with D-branes 
\footnote{
In this paper we will focus on the $\mathcal{A}_{4}$ algebra, 
however, the Nambu-Poisson 3-algebra 
and the Lorentzian 3-algebra have been proposed as 
the escapes from the restriction 
by relaxing the condition on dimensionality 
and the requirement of a positive definite metric respectively. 
}. 

%
%
%
%
%
%
%
%
%
%
%
%
%
%
%
%
%
%

\subsection{ABJM-model}
\label{secm2b}
The ABJM-model is a three-dimensional $\mathcal{N}=6$ superconformal 
$U(N)_{k}\times \hat{U}(N)_{-k}$ Chern-Simons-matter theory 
proposed as a generalization of the BLG-model in that 
it may describe the dynamics of an arbitrary number of coincident M2-branes 
\cite{Aharony:2008ug}. 
The theory has manifestly only $\mathcal{N}=6$ 
supersymmetry and the corresponding $SU(4)_{R}$ R-symmetry 
at the classical level. 
It has been discussed that
\cite{Aharony:2008ug,Gustavsson:2009pm,Kwon:2009ar} 
at $k=1$ and $k=2$ 
these symmetries are enhanced to $\mathcal{N}=8$ supersymmetry 
and $SO(8)_{R}$ R-symmetry as a quantum effect.

The theory contains 
four complex scalar fields $Y^{A}$, 
four complex spinors $\psi_{A}$ 
and two different types of gauge fields $A_{\mu}$ and $\hat{A}_{\mu}$. 
Here the upper and lower indices $A,B,\cdots=1,2,3,4$ denote 
$\bm{4}$ and $\overline{\bm{4}}$ of the $SU(4)_{R}$ respectively. 
The matter fields are $N\times N$ matrices so that 
$Y^{A}$ and $\psi_{A}$ transform as
$(\bm{N},\overline{\bm{N}})$ bi-fundamental representations 
of $U(N)_{k}\times \hat{U}(N)_{-k}$ gauge group, 
while $Y_{A}^{\dag}$ and $\psi^{\dag A}$ do as
$(\overline{\bm{N}},\bm{N})$. 
$A_{\mu}$ is a Chern-Simons $U(N)$ gauge field 
of level $+k$ and $\hat{A}_{\mu}$ is that of level $-k$. 
Also in the theory 
there is a $U(1)_{B}$ flavor symmetry 
and the corresponding baryonic charges 
are assigned $+1$ for bi-fundamental fields, 
$-1$ for anti-bi-fundamental fields 
and $0$ for gauge fields. 
%
%

The Lagrangian of the ABJM-model is given by 
\cite{
Benna:2008zy}
\begin{align}
\label{abjmlag1}
 \mathcal{L}_{\textrm{ABJM}}
=&-\textrm{Tr}
(D_{\mu}Y_{A}^{\dag}D^{\mu}Y^{A})
-i\textrm{Tr}(\psi^{\dag A}
\gamma^{\mu}D_{\mu}\psi_{A})
-V_{\textrm{ferm}}-V_{\textrm{bos}}\nonumber\\
&+\frac{k}{4\pi}\epsilon^{\mu\nu\lambda}\textrm{Tr}\left[
A_{\mu}\partial_{\nu}A_{\lambda}+\frac{2i}{3}
A_{\mu}A_{\nu}A_{\lambda}
-\hat{A}_{\mu}\partial_{\nu}\hat{A}_{\lambda}
-\frac{2i}{3}\hat{A}_{\mu}\hat{A}_{\nu}\hat{A}_{\lambda}\right]
\end{align}
where
\begin{align}
\label{abjmv01}
 V_{\textrm{ferm}}
=&-\frac{2\pi i}{k}
\textrm{Tr}
\Bigl(Y_{A}^{\dag}Y^{A}\psi^{\dag B}\psi_{B}
-\psi^{\dag B}Y^{A}Y_{A}^{\dag}\psi_{B}\nonumber\\
&-2Y_{A}^{\dag}Y^{B}\psi^{\dag A}\psi_{B}
+2Y^{A}Y_{B}^{\dag}\psi_{A}\psi^{\dag B}\nonumber\\
&-\epsilon^{ABCD}Y_{A}^{\dag}
\psi_{B}Y_{C}^{\dag}\psi_{D}
+\epsilon_{ABCD}Y^{A}\psi^{\dag B}Y^{C}\psi^{\dag D}\Bigr)\\
\label{abjmv02}
 V_{\textrm{bos}}
=&-\frac{4\pi^{2}}{3k^{2}}
\textrm{Tr}
\Bigl( Y^{A}Y_{A}^{\dag}Y^{B}Y_{B}^{\dag}Y^{C}Y_{C}^{\dag}
+Y_{A}^{\dag}Y^{A}Y_{B}^{\dag}Y^{B}Y_{C}^{\dag}Y^{C}
\nonumber\\
&+4Y^{A}Y_{B}^{\dag}Y^{C}Y_{A}^{\dag}Y^{B}Y_{C}^{\dag}
-6Y^{A}Y_{B}^{\dag}Y^{B}Y_{A}^{\dag}Y^{C}Y_{C}^{\dag}\Bigr).
\end{align}
Here we use the Dirac matrix  
${(\gamma^{\mu})_{\alpha}}^{\beta}=(i\sigma_{2},\sigma_{1},\sigma_{3})$. 
The spinor indices are raised, 
$\theta^{\alpha}=\epsilon^{\alpha\beta}\theta_{\beta}$, 
and lowered, $\theta_{\alpha}=\epsilon_{\alpha\beta}\theta^{\beta}$ 
with $\epsilon^{12}=-\epsilon_{12}=1$. 
Note that this makes the Dirac matrix 
$\gamma^{\mu}_{\alpha\beta}:=
{(\gamma^{\mu})_{\alpha}}^{\gamma}\epsilon_{\beta\gamma}
=(-\mathbb{I}_{2},-\sigma_{3},\sigma_{1})$ 
symmetric and guarantees the Hermiticity 
of the fermionic kinetic term. 
The covariant derivatives are defined by
\begin{align}
\label{abjmcov1}
 D_{\mu}Y^{A}&=\partial_{\mu}Y^{A}+iA_{\mu}Y^{A}-iY^{A}\hat{A}_{\mu},&
D_{\mu}\psi_{A}&=\partial_{\mu}\psi_{A}+iA_{\mu}\psi_{A}-i\psi_{A}\hat{A}_{\mu}
\nonumber\\
 D_{\mu}Y^{\dag}_{A}&=\partial_{\mu}Y^{\dag}_{A}-iA_{\mu}Y^{\dag}_{A}
+iY_{A}^{\dag}\hat{A}_{\mu},&
D_{\mu}\psi^{\dag A}&=\partial_{\mu}\psi^{\dag A}-iA_{\mu}\psi^{\dag A}
+i\psi^{\dag A}\hat{A}_{\mu}.
\end{align}

The supersymmetry transformation laws are
\begin{align}
\label{abjms01}
 \delta Y^{A}
&=i\omega^{AB}\psi_{B}\\
\label{abjms02}
\delta Y^{\dag}_{A}
&=i\psi^{\dag B}\omega_{AB}\\
\label{abjms03}
 \delta \psi_{A}
&=-\gamma^{\mu}\omega_{AB}D_{\mu}Y^{B}
+\frac{2\pi}{k}\left[
-\omega_{AB}(Y^{C}Y_{C}^{\dag}Y^{B}-Y^{B}Y_{C}^{\dag}Y^{C})
+2\omega_{CD}Y^{C}Y_{A}^{\dag}Y^{D}
\right]\\
\label{abjms04}
 \delta\psi^{\dag A}
&=D_{\mu}Y_{B}^{\dag}\omega^{AB}\gamma^{\mu}
+\frac{2\pi}{k}\left[
-(Y_{B}Y^{C}Y_{C}^{\dag}-Y_{C}^{\dag}Y^{C}Y_{B}^{\dag})\omega^{AB}
+2Y_{D}^{\dag}Y^{A}Y_{C}^{\dag}\omega^{CD}
\right]\\
\label{abjms05}
\delta A_{\mu}
&=\frac{\pi}{k}
\left(
-Y^{A}\psi^{\dag B}\gamma_{\mu}\omega_{AB}
+\omega^{AB}\gamma_{\mu}\psi_{A}Y_{B}^{\dag}
\right)\\
\label{abjms06}
\delta\hat{A}_{\mu}
&=\frac{\pi}{k}
\left(
-\psi^{\dag A}Y^{B}\gamma_{\mu}\omega_{AB}
+\omega^{AB}\gamma_{\mu}Y_{A}^{\dag}\psi_{B}
\right).
\end{align}
The parameter $\omega_{AB}$ is defined by
\begin{align}
\label{abjmsusypara1}
\omega_{AB}:=\epsilon_{i}(\Gamma^{i})_{AB},\ \ \ \ 
\omega^{AB}:=\epsilon_{i}(\Gamma^{i*})^{AB}
\end{align}
where the $SL(2,\mathbb{R})$ spinor $\epsilon^{i}$, $i=1,\cdots,6$ 
transforms as the representation $\bm{6}$ under the $SU(4)_{R}$ and 
$\Gamma^{i}$ is the six-dimensional $4\times 4$ matrix satisfying
\begin{align}
\label{abjmsp1}
(\Gamma^{i})_{AB}&=-(\Gamma^{i})_{BA}\\
\frac12 \epsilon^{ABCD}(\Gamma^{i})_{CD}&=-(\Gamma^{i\dag})^{AB}
=(\Gamma^{i*})^{AB}\\
\left\{\Gamma^{i},\Gamma^{j}\right\}&=2\delta_{ij}.
\end{align}
Note that the supersymmetry parameter $\omega_{AB}$ obeys
\begin{align}
\omega^{AB}&=\omega_{AB}^{*}=\frac12 \epsilon^{ABCD}\omega_{CD}.
\end{align}

The moduli space of the 
$U(N)_{k}\times \hat{U}(N)_{-k}$ ABJM-model is \cite{Aharony:2008ug} 
\begin{align}
\label{abjmmoduli0}
\mathcal{M}_{N,k}
=\frac{(\mathbb{C}^{4}/\mathbb{Z}_{k})^{N}}{S_{N}}
=\mathrm{Sym}^{N}(\mathbb{C}^{4}/\mathbb{Z}_{k}).
\end{align}
This can be identified with the moduli space 
of $N$ indistinguishable M2-branes moving 
in $\mathbb{C}^{4}/\mathbb{Z}_{k}$ transverse space. 
Therefore the ABJM-model is expected 
to describe the low-energy world-volume theory 
of $N$ coincident M2-branes probing an orbifold 
$\mathbb{C}^{4}/\mathbb{Z}_{k}$. 
The four complex scalar fields $Y^{A}$ 
represent the positions of the membranes in $\mathbb{C}^{4}$. 

In \cite{Lambert:2010ji} 
it has been discussed that 
if $N$ and $k$ are co-prime, 
then the vacuum moduli space of the $U(N)_{k}\times \hat{U}(N)_{-k}$ theory
is equivalent to that of the $SU(N)\times SU(N)/\mathbb{Z}_{N}$ theory. 
Consequently there are conjectural dualities 
between the ABJM theory and the BLG theory
\begin{align}
\label{abjmblg}
U(2)_{1}\times \hat{U}(2)_{-1}\ 
\textrm{ABJM theory}
&\Leftrightarrow 
SO(4)\ 
\textrm{BLG theory with}\ 
k=1\\
U(2)_{2}\times \hat{U}(2)_{-2}\ 
\textrm{ABJM theory}
&\Leftrightarrow
Spin(4)\ 
\textrm{BLG theory with}\ 
k=2. 
\end{align} 
These proposed dualities have been tested 
by the computations of the superconformal indices
\cite{Bashkirov:2011pt}. 
Hence we may regard the $SO(4)$ BLG-model with $k=1$ 
as the world-volume theory of two planar M2-branes 
propagating in a flat space.

\section{SCQM from flat M2-branes}
\label{secflat}
\subsection{$\mathcal{N}=16$ superconformal mechanics}
\label{secflat1}
\subsubsection{Derivation of quantum mechanics}
We begin our discussion with the BLG-model 
in the case where 
the membranes wrap a torus $T^{2}$ 
and propagate in a transverse space with an $SO(8)$ holonomy group. 
In this case the world-volume theory of M2-branes is given by 
the action (\ref{blglagrangian}) defined 
on $M_{3}=\mathbb{R}\times T^{2}$. 

In general a torus can be characterized by 
two periods in the complex plane. 
Such periods are defined as the integration 
of a holomorphic differential $\omega$ 
along two canonical homology basis $a$, $b$ of a torus. 
Let us define the periods by
\begin{align}
\label{period0}
\int_{a}\omega=1,\ \ \ \ \ 
\int_{b}\omega=\tau
\end{align}
where $\tau$ is the moduli of the torus and 
it should not be real. 

In the following we want to consider the limit in which $T^{2}$ has 
vanishingly small size and derive the low-energy effective 
one-dimensional theory on $\mathbb{R}$. 
In order to obtain such a theory 
we need to determine the configurations with the lowest energy. 
Since we are now considering supersymmetric theories, 
the conditions are expressed as the BPS equations. 
As we are interested in bosonic BPS configurations, 
we require that 
the background values of the fermionic fields vanish. 
Then the bosonic fields are automatically invariant under 
their supersymmetry transformations. 
Therefore the BPS equations correspond to the vanishing of the 
supersymmetry transformations (\ref{blgsusy2}) for fermionic fields. 
Also we discard the terms which include the covariant derivatives 
with respect to time because we are now interested in 
the low energy dynamics as a fluctuation around 
gauge invariant static configurations. 
Then one finds the BPS equations
\begin{align}
\label{bpst2a}
&D_{z}X_{a}^{I}=0,\ \ \ 
 D_{\overline{z}}X_{a}^{I}=0\\
\label{bpst2b}
&[X^{I},X^{J},X^{K}]=0.
\end{align}

To go further we consider 
the $SO(4)$ BLG-model that may describe two M2-branes. 
In this case the Higgs fields transform as fundamental representations 
of the $SO(4)$ gauge group and we assume that these Higgs fields 
have non-zero values. 
Then the generic solution to (\ref{bpst2b}) is given by 
$X_{a}^{I}=\left(X^{I}_{1},X^{I}_{2},0,0\right)^{T}$. 
For these solutions, the remaining BPS equations (\ref{bpst2a}) reduce to 
\begin{align}
\label{bps12a1}
\partial_{z}X_{1}^{I}+\tilde{A}_{z2}^{1}X_{2}^{I}&=0,& 
\partial_{z}X_{2}^{I}-\tilde{A}_{z2}^{1}X_{1}^{I}&=0\\
\label{bps12b1}
\tilde{A}_{z3}^{1}X_{1}^{I}+\tilde{A}_{z3}^{2}X_{2}^{I}&=0,&
\tilde{A}_{z4}^{1}X_{1}^{I}+\tilde{A}_{z4}^{2}X_{2}^{I}&=0
\end{align}
and their complex conjugates. 
First of all, the equations (\ref{bps12a1}) tell us that 
the sum of the squares 
$(X_{1}^{I})^{2}+(X_{2}^{I})^{2}$ for 
$I=1,\cdots,8$ is independent 
of the locus of the Riemann surface.  
Thus we can write 
\begin{align}
\label{bosconf1a}
X_{1}^{I+2}+iX_{2}^{I+2}
=r^{I}
e^{i(\theta^{I}+\varphi(z,\overline{z}))}
\end{align}
where $r^{I}, \theta^{I} \in \mathbb{R}$ are constant on the torus  
and represent the configuration of the two membranes in the $I$-th
direction while 
$\varphi(z,\overline{z})$ may depend on $z$ and $\overline{z}$. 
Furthermore the equations (\ref{bps12a1}) 
enable us to write $\tilde{A}_{z2}^{1}=\partial_{z}\varphi$. 
The second set of equations (\ref{bps12b1}) 
forces us to turn off four of six gauge fields; 
$\tilde{A}_{z3}^{1}=\tilde{A}_{z3}^{2}=
\tilde{A}_{z4}^{1}=\tilde{A}_{z4}^{2}=0$. 
These components of the gauge field become massive by the Higgs mechanism. 
Note that the above set of solutions automatically satisfies 
the integrability condition for (\ref{bpst2a}) 
because the gauge field $\tilde{A}_{z2}^{1}$ is flat.

One can find further restrictions by noting that the flat gauge fields 
$\tilde{A}_{z2}^{1}$ on a torus have specific expressions. 
Cutting a torus along the canonical basis $a$ and $b$, 
the sections of a flat bundle are 
described by their transition functions, 
i.e. constant phases around $a$ and $b$. 
Thus they can be completely classified by their twists 
$e^{2\pi i\xi}$, $e^{-2\pi i\zeta}$ 
on the homology along cycles $a$, $b$ 
where $\xi$ and $\zeta$ are real parameters.  
This space is the torus $\mathbb{C}/L_{\tau}$ 
where $L_{\tau}$ is the lattice generated by 
$\mathbb{Z}+\tau\mathbb{Z}$. 
It is referred to as the Jacobi variety of $T^{2}$ denoted
by $\mathrm{Jac}(T^{2})$. 
The twists on the homology can be described as a point 
on the Jacobi variety. 
Hence the flat gauge field can be expressed in the form 
\cite{AlvarezGaume:1986es}
\begin{align}
\label{az12}
\tilde{A}_{z2}^{1}
=-2\pi\frac{\Theta}{\tau-\overline{\tau}}\omega,\ \ \ \ \ 
\tilde{A}_{\overline{z}2}^{1}
=2\pi\frac{\overline{\Theta}}{\tau-\overline{\tau}}
\overline{\omega}
\end{align}
where $\Theta:=\zeta+\overline{\tau}\xi$ is   
the complex parameter representing the twists on 
the homology along two cycles.  
Subsequently we can write 
\begin{align}
\label{phi}
\varphi(z,\overline{z})
=2\pi\frac{\overline{\Theta}}{\tau-\overline{\tau}}\overline{z}
-2\pi\frac{\Theta}{\tau-\overline{\tau}}z. 
\end{align}
Recalling that 
the angular variable $\varphi(z,\overline{z})$ in the 
$X^{I}_{1} X_{2}^{I}$-plane 
characterizes the ratio of two bosonic degrees of freedom 
for the two membranes, 
it must take same values modulo $2\pi \mathbb{Z}$ under 
the shifts $z\rightarrow z+1$ and $z\rightarrow z+\tau$ around two cycles. 
This implies that both the coordinates $\xi$ and $\zeta$ 
can only have integer values, namely $\tilde{A}_{z2}^{1}$ 
and $\tilde{A}_{\overline{z}2}^{1}$ are quantized. 
Therefore the generic BPS solutions are given by
\begin{align}
\label{0bos2}
X^{I+2}
&=
\left(
\begin{array}{c}
X_{A}^{I}\\
X_{B}^{I}\\
0\\
0\\
\end{array}
\right)
=\left(
\begin{array}{c}
\cos(\theta^{I}+\varphi(z,\overline{z}))\\
\sin(\theta^{I}+\varphi(z,\overline{z}))\\
0\\
0\\
\end{array}
\right)r^{I}
\nonumber\\
\tilde{A}_{z}
&=\left(
\begin{array}{cccc}
0&-2\pi\frac{\Theta}{\tau-\overline{\tau}}\omega_{z}&0&0\\
2\pi\frac{\Theta}{\tau-\overline{\tau}}\omega_{z}&0&0&0\\
0&0&0&\tilde{A}_{z4}^{3}(z,\overline{z})\\
0&0&-\tilde{A}_{z4}^{3}(z,\overline{z})&0\\
\end{array}
\right).
\end{align}
Here $\tilde{A}_{z4}^{3}$ and $\tilde{A}_{\overline{z}4}^{3}$ 
are the Abelian gauge fields associated 
with the preserved $U(1)$ symmetry 
and have no constraints from the BPS conditions. 
Taking into account the bosonic configurations (\ref{0bos2}) 
and the supersymmetry transformations (\ref{blgsusy1}), 
we introduce fermionic partners  
\begin{align}
\label{0fer1}
\Psi_{\pm}
=\left(
\begin{array}{c}
\Psi_{\pm A}\\
\Psi_{\pm B}\\
0\\
0\\
\end{array}
\right),\ \ \ 
\overline{\Psi}^{\pm}=\left(
\begin{array}{c}
\overline{\Psi}^{\pm}_{A}\\
\overline{\Psi}^{\pm}_{B}\\
0\\
0\\
\end{array}
\right)
\end{align}
where 
$\overline{\Psi}$ is the conjugate spinor defined by 
$\overline{\Psi}:=\Psi^{T}\tilde{C}$ in terms of 
the $SO(8)$ charge conjugation matrix $\tilde{C}$. 
$\Psi_{+}^{a}$ and $\overline{\Psi}^{+a}$ 
are the $SO(2)_{E}$ spinors with the positive chiralities 
while $\Psi_{-}^{a}$ and $\overline{\Psi}^{-a}$ 
carry the negative ones.  
Both of them transform as $\bm{8}_{c}$ of the $SO(8)_{R}$.

Given the above static BPS configurations (\ref{0bos2}) and (\ref{0fer1}), 
we now wish to consider the evolution of time 
and compactify the system on $T^{2}$. 
Substitution of the configurations (\ref{0bos2}) and (\ref{0fer1}) 
into the action (\ref{blglagrangian}) yields
\begin{align}
\label{effs0}
S=
\int_{\mathbb{R}}dt
\int_{T^{2}}d^{2}z 
&
\Biggl[
\frac12 D_{0}X^{Ia}D_{0}X_{a}^{I}
-\frac{i}{2}
\overline{\Psi}^{\alpha a}D_{0}\Psi_{\alpha a}\nonumber\\
&-\frac{k}{2\pi}\tilde{A}_{02}^{1}\tilde{F}_{z\overline{z}4}^{3}
-\frac{k}{4\pi}
\left(
\tilde{A}_{z2}^{1}\dot{\tilde{A}}_{\overline{z}4}^{3}
-\tilde{A}_{\overline{z}2}^{1}\dot{\tilde{A}}_{z4}^{3}
\right)
\Biggr]
\end{align}
where the Greek letters $\alpha=+,-$ 
denote the $SO(2)_{E}$ spinor indices. 
The terms in the first line of the action (\ref{effs0}) 
come from the kinetic terms of the BLG action 
while those in the second correspond 
to the twisted topological Chern-Simons terms. 

Firstly since the gauge fields $\tilde{A}_{z2}^{1}$ and 
$\tilde{A}_{\overline{z}2}^{1}$ are quantized and 
their time derivatives do not appear in the action, 
these fields are just auxiliary fields. 
Exploiting the equations of motion they can be excluded 
and we find the constraints 
$\dot{\tilde{A}}_{z4}^{3}=\dot{\tilde{A}}_{\overline{z}4}^{3}=0$. 
Hence the corresponding field strength 
$\tilde{F}_{z\overline{z}4}^{3}$ has no time dependence. 
In order to dimensionally reduce the theory on the torus, 
we rescale the fields as
\begin{align}
\label{resc1}
X^{I'}=R^{2}X^{I}, \ \ \ \ \ 
\Psi_{\alpha a}'=R^{2}\Psi_{\alpha a}, \ \ \ \ \ 
{\overline{\Psi}^{\alpha a}}'=R^{2}\overline{\Psi}^{\alpha a}
\end{align}
where $R$ is the circumference of the torus. 
Note that they get the canonical dimensions in the reduced theory; 
the bosonic variable $X^{I'}$ has 
mass dimension $-1/2$ and 
the fermionic variable $\Psi'$ acquires mass dimension $0$.

Performing the integration on the torus 
by means of the Kaluza-Klein ansatz for $\tilde{A}_{0}^{12}$ 
and dropping the primes on the fields, 
one finds the effective action
\begin{align}
\label{effs1}
S=
\int_{\mathbb{R}}dt
\Biggl[
\frac12 D_{0}X^{Ia}D_{0}X^{I}_{a}
-\frac{i}{2}
\overline{\Psi}^{\alpha a}D_{0}\Psi_{\alpha a}
-kC_{1}(E)\tilde{A}_{02}^{1}
\Biggr].
\end{align}
Here 
\begin{align}
\label{ch01}
C_{1}(E)=\int_{T^{2}}c_{1}(E)
:=\frac{1}{2\pi}\int_{T^{2}}
d^{2}z \tilde{F}_{z\overline{z}4}^{3}
\end{align}
is the Chern number resulting from the integration 
of the first Chern class $c_{1}(E)$ of the 
$U(1)$ principal bundle $E\rightarrow T^{2}$ over the torus, 
which is associated with the preserved $U(1)$ gauge field
$\tilde{A}_{z4}^{3}$.

The action (\ref{effs1}) is invariant 
under the one-dimensional conformal transformations
\begin{align}
\label{tconf1}
\delta t&=f(t)=a+bt+ct^{2}, 
&\delta\partial_{0}&=-\dot{f}\partial_{0}\\
\delta X_{a}^{I}&=\frac12 \dot{f} X_{a}^{I}, 
&\delta \tilde{A}_{02}^{1}&=-\dot{f}\tilde{A}_{02}^{1}\\
\delta \Psi_{\alpha a}&=0,&
\delta \overline{\Psi}^{\alpha a}&=0
\end{align}
where $f(t)$ is a quadratic function of time with real 
infinitesimal parameters $a$, $b$ and $c$. 

Besides, the action (\ref{effs1}) is invariant 
under the $\mathcal{N}=16$ supersymmetry transformations
\begin{align}
\label{tsusy01}
\delta X^{I}_{a}
&=i\overline{\epsilon}^{+}\tilde{\Gamma}^{I}\Psi_{-a}
-i\overline{\epsilon}_{-}\tilde{\Gamma}^{I}\Psi_{+a}, 
&
\delta \tilde{A}_{02}^{1}&=0
\\
\label{tsusy02}
\delta \Psi_{+a}
&=-D_{0}X_{a}^{I}\tilde{\Gamma}^{I}\epsilon_{-}, 
&\delta \Psi_{-a}
&=D_{0}X_{a}^{I}\tilde{\Gamma}^{I}\epsilon_{+}. 
\end{align}
Therefore the resulting effective theory (\ref{effs1}) takes the form 
of $\mathcal{N}=16$ superconformal gauged quantum mechanics with 
a Fayet-Iliopoulos (FI) term.

%
%
%
%
%
%
%
%

\subsubsection{Reduced system with inverse-square interaction}
\label{secflat2}
Since the gauged mechanical action (\ref{effs1}) 
is quadratic in the $U(1)$ gauge field $\tilde{A}_{02}^{1}$ 
and does not involve the time derivative of it, 
$\tilde{A}_{02}^{1}$ 
is identified with an auxiliary field 
and has no contribution to the Hamiltonian. 
Hence the Hamiltonian is invariant under the action 
of the corresponding $U(1)$ gauge group on the phase space $\mathcal{M}$.  
This means that 
the corresponding moment map $\mu:\mathcal{M}\rightarrow \mathfrak{u}(1)^{*}$ 
is the integral of motion \cite{MR0690288} and 
one can reduce the given phase space $\mathcal{M}$ 
to a smaller one $\mathcal{M}_{c}=\mu^{-1}(c)$ 
with fewer degrees of freedom by fixing the inverse of 
the moment map at a point $c\in \mathfrak{u}(1)^{*}$ 
\footnote{
The components of the moment map form a system 
being in involution since the gauge group is Abelian. 
So we do not need to divide by the 
non-trivial coadjoint isotropy subgroup to obtain the reduced phase space.}.

In fact it is known that one-dimensional gauged matrix models give rise to 
the alternative  descriptions of the Calogero model and 
its generalizations as the reduced systems 
\cite{MR0478225,Polychronakos:1991bx,Fedoruk:2008hk}. 
In order to obtain our reduced system, 
we shall eliminate the auxiliary field $\tilde{A}_{02}^{1}$ in two steps; 
first we choose a specific gauge 
and then impose the Gauss law constraint 
to ensure the consistency of the gauge fixing. 
Let us choose the temporal gauge $\tilde{A}_{0}=0$. 
Together with the solutions
\begin{align}
\label{0a0}
\tilde{A}_{02}^{1}
&=\frac{kC_{1}(E)+\sum_{I}(r^{I})^{2}\dot{\theta}^{I}
+i\overline{\Psi}^{\alpha}_{A}\Psi_{\alpha B}
}
{\sum_{I}(r^{I})^{2}}\\
\tilde{A}_{03}^{1}
&=\tilde{A}_{04}^{1}
=\tilde{A}_{03}^{2}
=\tilde{A}_{04}^{2}=0
\end{align}
to the equations of motion for $\tilde{A}_{0}$, 
we can read off the Gauss law constraint
\begin{align}
\label{1a0}
\phi_{0}:=kC_{1}(E)+
\sum_{I}(r^{I})^{2}\dot{\theta}^{I}
+i\overline{\Psi}^{\alpha}_{A}\Psi_{\alpha B}=0.
\end{align}
This equation is the moment map condition. 
To see the physical meaning of this constraint, 
we observe that $(r^{I})^{2}\dot{\theta}^{I}$ 
represents the ``angular momentum'', the $SO(2)$-charge corresponding to
the rotation in the $X^{I}_{1}X^{I}_{2}$-plane 
while the fermionic bilinear term 
$i\overline{\Psi}_{A}^{\alpha}\Psi_{\alpha B}$ 
produces the charge of the $SO(2)$ rotational group of 
the two types of fermionic variables $\Psi_{A}$ and $\Psi_{B}$.  
Accordingly the equation (\ref{1a0}) says that 
the total $SO(2)$ charge which rotates the internal degrees of freedom 
for the two M2-branes is fixed by the Chern-Simons level $k$ 
and the Chern number $C_{1}(E)$.

With the constraint function $\phi_{0}$, 
one can write a new Lagrangian by adding $\lambda \phi_{0}$ 
where $\lambda$ is the Lagrange multiplier. 
The resulting action is
\begin{align}
\label{effs2}
S=\int_{\mathbb{R}}
dt &\Biggl[
\frac12 \sum_{I}(\dot{r}^{I})^{2}
+\frac12 \sum_{I} (r^{I}\dot{\theta}^{I})^{2}
-\frac{i}{2}\overline{\Psi}^{\alpha a}\dot{\Psi}_{\alpha a}\nonumber\\
&+\lambda\left(
kC_{1}(E)+
\sum_{I}(r^{I})^{2}\dot{\theta}^{I}
+i\overline{\Psi}_{A}^{\alpha}\Psi_{\alpha B}
\right)
\Biggr].
\end{align}
The absence of the variables $\theta^{I}$'s in the action (\ref{effs2}) 
immediately implies that they are cyclic coordinates and 
their canonical momenta $p_{\theta^{I}}=(r^{I})^{2}\dot{\theta}^{I}$ 
are just the integrals of motion. 

It is possible to eliminate cyclic coordinates 
from the Lagrangian by constructing a new Lagrangian, 
the so-called Routhian. 
The Routhian is a hybrid 
between the Lagrangian and the Hamiltonian, 
defined by performing a Legendre transformation 
on the cyclic coordinates
\begin{align}
\label{routh1}
R(r^{I},\dot{r}^{I},h^{I},\Psi):=
L(r^{I},\dot{r}^{I},\dot{\theta}^{I},\Psi)
-\sum_{I}\dot{\theta}^{I}p_{\theta^{I}}.
\end{align}
Due to the partial Legendre transformation, 
the variables $r^{I}$ and $\Psi$ still follow the Euler-Lagrange equations 
while the ignorable coordinates $\theta^{I}$ and their momenta 
$h^{I}:=p_{\theta^{I}}$ 
obey the Hamilton equations. 
However, the latter set of equations results in trivial statements; 
the constant property of $h^{I}$ (i.e. $\dot{h}^{I}=0$) 
and the definition of $h^{I}$ 
(i.e. $\dot{\theta}^{I}=\frac{h^{I}}{(r^{I})^{2}}$). 
So classically the Routhian is not really 
$R(r^{I},\dot{r}^{I}, h^{I},\Psi)$ 
but rather $R(r^{I},\dot{r}^{I},\Psi)$ 
along with the integrals of motion $h^{I}$'s. 
Hence we can  rewrite (\ref{effs2}) as
\begin{align}
\label{effs3}
S=\int_{\mathbb{R}}
dt &\Biggl[ 
\frac12 \sum_{I}(\dot{r}^{I})^{2}
-\frac12\sum_{I}\frac{(h^{I})^{2}}{(r^{I})^{2}}
-\frac{i}{2}\overline{\Psi}^{\alpha a}\dot{\Psi}_{\alpha a}
+\lambda\left(
kC_{1}(E)+\sum_{I}h^{I}
+i\overline{\Psi}^{\alpha}_{A}\Psi_{\alpha B}
\right)
\Biggr].
\end{align}
Integrating out $\lambda$, 
we finally obtain the reduced effective action 
\begin{align}
\label{0eff0}
S=\frac12\ \int_{\mathbb{R}}
dt 
\Biggl[
&\dot{q}^{2}
+\sum_{I\neq K}(\dot{r}^{I})^{2}
-i\overline{\Psi}^{\alpha a}\dot{\Psi}_{\alpha a}\nonumber\\
&-\frac{\left[
kC_{1}(E)+\sum_{I\neq K}h^{I}
+i\overline{\Psi}^{\alpha}_{A}\Psi_{\alpha B}\right]^{2}}{q^{2}}
-\sum_{I\neq K}\frac{(h^{I})^{2}}{(r^{I})^{2}}
\Biggr].
\end{align}
Here we have defined $q:=r^{K}$ where $K$ denotes the specific direction 
in which $h^{K}$ is automatically determined by other conserved
quantities $h^{I}$'s. 
Note that the terms appearing in the numerator of the potential 
are the integrals of motion, 
namely they commute with the Hamiltonian.

Let us study the classical properties of the theory (\ref{0eff0}). 
The action (\ref{0eff0}) leads to the classical equations of motion
\begin{align}
\label{eomqm1}
\ddot{q}&=\frac{[kC_{1}(E)+\sum_{I\neq K}h^{I}
+i\overline{\Psi}^{\alpha}_{A}\Psi_{\alpha B}]^{2}}{q^{3}}\\
\label{eomqm1a}
\ddot{r}^{I}&=\frac{(h^{I})^{2}}{(r^{I})^{3}}\\
\label{eomqm2a}
\dot{\Psi}_{\alpha A}
&=-\frac{[kC_{1}(E)+\sum_{I\neq K}h^{I}
+i\overline{\Psi}^{\alpha}_{A}\Psi_{\alpha B}]}{q^{2}}
\Psi_{\alpha B}\\
\label{eomqm3a}
\dot{\Psi}_{\alpha B}
&=\frac{[kC_{1}(E)+\sum_{I\neq K}h^{I}
+i\overline{\Psi}_{A}^{\alpha}\Psi_{\alpha B}]}{q^{2}}
\Psi_{\alpha A}. 
\end{align}
Making use of the equations of motion 
(\ref{eomqm2a}) and (\ref{eomqm3a}), 
one can check that the differentiation 
of the Gauss law constraint 
(\ref{1a0}) with respect to time $t$ vanishes. 
In other words, $\phi_{0}$ is the constant of motion.

The canonical momenta are
\begin{align}
\label{mom01}
p&:=\frac{\partial L}{\partial \dot{q}}=\dot{q},&
p_{I}&:=\frac{\partial L}{\partial \dot{r}^{I}}=\dot{r}^{I}\\
\label{mom02}
\pi^{\alpha a}
&:=\frac{\vec{\partial}L}
{\partial \dot{\Psi}_{\alpha a}}
=\frac{i}{2}\overline{\Psi}^{\alpha a}, 
&\overline{\pi}_{\alpha a}
&:=\frac{\vec{\partial} L}
{\partial \dot{\overline{\Psi}}^{\alpha a}}
=\frac{i}{2}\Psi_{\alpha a}. 
\end{align}
The fermionic momenta $\pi^{\alpha a}$ and $\overline{\pi}_{\alpha a}$ 
do not depend on the velocities 
but on the fermionic degrees of freedom themselves. 
Hence one can read second-class constraints
\begin{align}
\label{prcons1}
\phi_{1}^{\alpha a}
&:=\pi^{\alpha a}-\frac{i}{2}\overline{\Psi}^{\alpha a}=0,&
\phi_{2\alpha a}
&:=\overline{\pi}_{\alpha a}-\frac{i}{2}\Psi_{\alpha a}=0.
\end{align}
Under the constraints, 
we get the Dirac brackets
\begin{align}
\label{dirac1}
\left[q,p\right]_{DB}&=1,& 
\left[r^{I},p_{J}\right]_{DB}&={\delta^{I}}_{J}\\ 
\label{dirac2}
\left[
\Psi_{\alpha a\dot{A}},\pi^{\beta b\dot{B}}
\right]_{DB}&=
-\frac12 \delta_{\alpha\beta}\delta_{ab}\delta_{\dot{A}\dot{B}},&
\left[
\Psi_{\alpha a\dot{A}},\overline{\Psi}^{\beta b\dot{B}}
\right]_{DB}
&=i\delta_{\alpha\beta}\delta_{ab}\delta_{\dot{A}\dot{B}}.
\end{align}

The action (\ref{0eff0}) 
is invariant under the following 
one-dimensional conformal transformations
\begin{align}
\label{conf0}
\delta t&=f(t)=a+bt+ct^{2},
&\delta\partial_{0}&=-\dot{f}\partial_{0}\\
\label{conf1} 
\delta q&=\frac12 \dot{f}q,
&\delta r^{I}&=\frac12 \dot{f}r^{I}\\
\label{conf2}
\delta \Psi_{\alpha a}&=0, 
&\delta \overline{\Psi}^{\alpha a}&=0.
\end{align}
Here the constant parameters $a$, $b$ and $c$ are infinitesimal parameters of  
translation, dilatation and conformal boost respectively. 
The corresponding Noether charges, 
the Hamiltonian $H$, the dilatation operator $D$ and 
the conformal boost operator $K$ are found to be
\begin{align}
\label{hkdconf1}
H&=\frac12 \left[
p^{2}+\frac{
\left(
kC_{1}(E)+\sum_{I\neq K}h^{I}
+i\overline{\Psi}^{\alpha}_{A}\Psi_{\alpha B}
\right)
^{2}}
{q^{2}}
+\sum_{I\neq K}\left(
p_{I}^{2}+\frac{(h^{I})^{2}}{(r^{I})^{2}}
\right)
\right]
\\
\label{hkdconf2}
D&=tH
-\frac14 
\left[\left(qp+pq\right)+\sum_{I\neq K}
\left( r^{I}p_{I}+p_{I}r^{I} \right)
\right]
\\
\label{hkdconf3}
K&=t^{2}H
-\frac{1}{2}t\left[
\left(qp+pq\right)
+\sum_{I\neq K}\left(
r^{I}p_{I}+p_{I}r^{I}
\right)
\right]
+\frac12 
\left[
q^{2}+\sum_{I\neq K}(r^{I})^{2}
\right].
\end{align}
The operators $D$ and $K$ 
are the constants of motion in the sense that 
\begin{align}
\label{dkconst1}
\frac{\partial D}{\partial t}+[H,D]_{DB}&=0,&
\frac{\partial K}{\partial t}+[H,K]_{DB}&=0.
\end{align}
Note that the explicit time dependence of $D$ and $K$ 
can be absorbed into the similarity transformations
\begin{align}
\label{simdk0}
D&=e^{itH}D_{0}e^{-itH},&
K&=e^{itH}K_{0}e^{-itH}
\end{align}
where
\begin{align}
\label{hkdconf2a}
D_{0}&:=-\frac14 
\left[\left(qp+pq\right)+\sum_{I\neq K}
\left(r^{I}p_{I}+p_{I}r^{I}\right)
\right]\\
\label{hkdconf3a}
K_{0}&:=\frac12 
\left[
q^{2}+\sum_{I\neq K}(r^{I})^{2}
\right]
\end{align}
are time independent parts of $D$ and $K$ respectively. 
So we will use the time independent parts as the explicit 
expressions for $D$ and $K$ and drop off the subscripts.  


The action (\ref{0eff0}) is invariant 
under the following fermionic transformations
\begin{align}
\label{susy001a}
\delta q&=
\frac{i}{\sqrt{2}}
\left(\overline{\epsilon}^{-}\Psi_{-A}
-\overline{\epsilon}^{+}\Psi_{+A}
\right)
+\frac{i}{\sqrt{2}}\left(
\overline{\epsilon}^{-}\Psi_{-B}
-\overline{\epsilon}^{+}\Psi_{+B}
\right)
\\
\label{susy001}
\delta r^{I}&=
i\cos\theta^{I}\left(
\overline{\epsilon}^{+}\tilde{\Gamma}^{I}\Psi_{-A}
-\overline{\epsilon}^{-}\tilde{\Gamma}^{I}\Psi_{+A}
\right)
+i\sin\theta^{I}\left(
\overline{\epsilon}^{+}\tilde{\Gamma}^{I}\Psi_{-B}
-\overline{\epsilon}^{-}\tilde{\Gamma}^{I}\Psi_{+B}
\right)
\\
\label{susy002}
\delta\Psi_{+A\dot{A}}
&=
-\frac{1}{\sqrt{2}}\left(\dot{q}-\frac{h^{K}}{q}\right)
\epsilon_{+\dot{A}}
-\frac{i}{\sqrt{2}}\frac{l}{q}\Psi_{+B\dot{A}}
-\sum_{I\neq K}\left(
\dot{r}^{I}\cos\theta^{I}-\sin\theta^{I}\frac{h^{I}}{r^{I}}
\right)\tilde{\Gamma}^{I}\epsilon_{-\dot{A}}\\
\label{susy003}
\delta\Psi_{-A\dot{A}}
&=
\frac{1}{\sqrt{2}}\left(\dot{q}-\frac{h^{K}}{q}\right)
\overline{\epsilon}_{-\dot{A}}
-\frac{i}{\sqrt{2}}\frac{l}{q}\Psi_{-B\dot{A}}
+\sum_{I\neq K}\left(
\dot{r}^{I}\cos\theta^{I}-\sin\theta^{I}\frac{h^{I}}{r^{I}}
\right)\tilde{\Gamma}^{I}\epsilon_{+\dot{A}}\\
\label{susy004}
\delta\Psi_{+B\dot{A}}
&=
-\frac{1}{\sqrt{2}}\left(\dot{q}+\frac{h^{K}}{q}\right)\epsilon_{+\dot{A}}
+\frac{i}{\sqrt{2}}\frac{l}{q}\Psi_{+A\dot{A}}
-\sum_{I\neq K}\left(
\dot{r}^{I}\sin\theta^{I}+\cos\theta^{I}\frac{h^{I}}{r^{I}}
\right)\tilde{\Gamma}^{I}\epsilon_{-\dot{A}}\\
\label{susy005}
\delta\Psi_{-B\dot{A}}
&=
\frac{1}{\sqrt{2}}\left(\dot{q}+\frac{h^{K}}{q}\right)\overline{\epsilon}_{-\dot{A}}
+\frac{i}{\sqrt{2}}\frac{l}{q}\Psi_{-A\dot{A}}
+\sum_{I\neq K}\left(
\dot{r}^{I}\sin\theta^{I}+\cos\theta^{I}\frac{h^{I}}{r^{I}}
\right)\tilde{\Gamma}^{I}\epsilon_{+\dot{A}}
\end{align}
where we have defined
\begin{align}
\label{theta01}
\theta^{I}(t)
&=h^{I}\int^{t} \frac{dt'}{(r^{I}(t'))^{2}}\\
\label{l01}
l&=
\left(
\overline{\Psi}^{+A}\epsilon_{+}
-\overline{\Psi}^{-A}\epsilon_{-}
\right)
-
\left(
\overline{\Psi}^{+B}\epsilon_{+}
-\overline{\Psi}^{-B}\epsilon_{-}
\right). 
\end{align}
We should note that the supersymmetry is generically non-local 
in the sense that the transformations contain the integrals of the function of 
the non-local variables $r^{I}$'s with respect to time. 
The non-locality is the consequence of the Routh reduction. 
Hence the infinite number of 
the associated conserved charges may exist 
and things may become much more exotic. 
However, as seen from the (\ref{hkdconf1}), 
the motion in the $K$-th direction endowed with  
the local supersymmetry and others with non-local ones
are essentially decoupled  
because their Hamiltonians commute with each other. 
Thus we can treat them separately. 
This indicates that the theory possesses 
the local conserved supercurrents 
and the non-local supercurrents which are in involution.

\subsubsection{$OSp(16|2)$ superconformal mechanics}
\label{secflat3}
Now we want to study the $K$-th motion 
associated with the local charges and shed light on the algebraic
structure of the symmetry group in the quantum mechanics. 
For simplicity let us consider the case where 
the all independent conserved charges $h^{I}$'s are zeros. 
This is realized when the internal degrees of freedom 
for two M2-branes are unbiased 
\footnote{This specific charge assignment does not affect the following discussion for the $K$-th motion 
since non-vanishing $h^{I}$'s can only give rise to  
a constant shift in the numerator of the potential.}.
The dynamics in the $K$-th direction is given by the action
\begin{align}
\label{effk1}
S=\frac12\ \int_{\mathbb{R}}
dt 
\Biggl[ 
\dot{q}^{2}
-i\overline{\Psi}^{\alpha a}\dot{\Psi}_{\alpha a}
-\frac{\left(
kC_{1}(E)
+i\overline{\Psi}^{\alpha}_{A}\Psi_{\alpha B}
\right)^{2}}
{q^{2}}
\Biggr].
\end{align}
The notable feature is that 
our reduced action (\ref{effk1}) has the predicted form 
for $\mathcal{N}>4$ superconformal mechanical action 
discussed in \cite{Ivanov:1988it,Wyllard:1999tm,Fedoruk:2011aa}. 
%

The action (\ref{effk1}) is invariant under 
the following $\mathcal{N}=16$ supersymmetry transformation laws
\begin{align}
\label{tsusy001}
\delta q&=
\frac{i}{\sqrt{2}}\left(
\overline{\epsilon}^{-}\Psi_{-A}
-\overline{\epsilon}^{+}\Psi_{+A}
\right)
+\frac{i}{\sqrt{2}}\left(
\overline{\epsilon}^{-}\Psi_{-B}
-\overline{\epsilon}^{+}\Psi_{+B}
\right)
\\
\label{tsusy002}
\delta \Psi_{+A\dot{A}}
&=
-\frac{1}{\sqrt{2}}\left(
\dot{q}+\frac{g}{q}
\right)\epsilon_{+\dot{A}}
-\frac{i}{\sqrt{2}}\frac{l}{q}\Psi_{+B\dot{A}}
\\
\label{tsusy003}
\delta \Psi_{-A\dot{A}}
&=
\frac{1}{\sqrt{2}}\left(\dot{q}+\frac{g}{q}\right)
\epsilon_{-\dot{A}}
-\frac{i}{\sqrt{2}}\frac{l}{q}\Psi_{-B\dot{A}}
\\
\label{tsusy004}
\delta\Psi_{+B\dot{A}}
&=
-\frac{1}{\sqrt{2}}\left(
\dot{q}-\frac{g}{q}
\right)\epsilon_{+\dot{A}}
+\frac{i}{\sqrt{2}}\frac{l}{q}\Psi_{+A\dot{A}}
\\
\label{tsusy005}
\delta \Psi_{-B\dot{A}}
&=
\frac{1}{\sqrt{2}}
\left(
\dot{q}-\frac{g}{q}
\right)\epsilon_{-\dot{A}}
+\frac{i}{\sqrt{2}}\frac{l}{g}\Psi_{-A\dot{A}}
\end{align}
where
\begin{align}
\label{g01}
g&=kC_{1}(E)
+i\overline{\Psi}_{A}^{\alpha}\Psi_{\alpha B}.
\end{align}
These supersymmetry transformations are local and
we therefore can apply the conventional Noether's procedure. 
By means of the Noether's method, 
the corresponding supercharges are calculated to be
\begin{align}
\label{tsch1}
Q_{+\dot{A}}
&=\frac{i}{\sqrt{2}}\left(p+\frac{g}{q}\right)\Psi_{+A\dot{A}}
+\frac{i}{\sqrt{2}}\left(p-\frac{g}{q}\right)\Psi_{+B\dot{A}}
\\
Q_{-\dot{A}}
&=
-\frac{i}{\sqrt{2}}\left(p+\frac{g}{q}\right)\Psi_{-A\dot{A}}
-\frac{i}{\sqrt{2}}\left(p-\frac{g}{q}\right)\Psi_{-B\dot{A}}.
\end{align}
Since the action (\ref{effk1}) is invariant 
under the conformal transformations 
$\delta t=f(t)$, $\delta q=\frac12 \dot{f}q$ and  
$\delta\Psi_{\alpha a}=0$, 
three generators, the Hamiltonian $H$, 
the dilatation generator $D$ and the conformal boost generator $K$ 
are explicitly expressed as 
\begin{align}
\label{hkdconf01}
H&=\frac12 p^{2}+\frac{\left[
kC_{1}(E)+i\overline{\Psi}^{\alpha}_{A}\Psi_{\alpha B}
\right]^{2}}{2q^{2}}\\
\label{hkdconf02}
D&=-\frac14 \{q,p\}\\
\label{hkdconf03}
K&=\frac12 q^{2}
\end{align}
where $\{ , \}$ represents an anti-commutator.

In order to quantize the theory, 
we impose the (anti)commutation relations 
for the canonical variables obtained from 
the Dirac brackets (\ref{dirac1}) and (\ref{dirac2}) 
\begin{align}
\label{canonical0}
[q,p]&=i,&
\left\{
\Psi_{\alpha a\dot{A}},\overline{\Psi}^{\beta b\dot{B}}
\right\}
&=-\delta_{\alpha\beta}\delta_{ab}\delta_{\dot{A}\dot{B}}.
\end{align}

The presence of the conformal symmetry and the supersymmetry 
leads to that of a superconformal symmetry. 
Let us define the superconformal boost generators
\begin{align}
\label{scqmboost01}
S_{+\dot{A}}&=\frac{i}{\sqrt{2}}q\left(
\Psi_{+A\dot{A}}+\Psi_{+B\dot{A}}
\right)\\
S_{-\dot{A}}&=-\frac{i}{\sqrt{2}}q\left(
\Psi_{-A\dot{A}}+\Psi_{-B\dot{A}}
\right).
\end{align}

Additionally the theory has the internal R-symmetry which rotates 
the fermionic charges. 
We define the R-symmetry generators by
\begin{align}
\label{j01}
(J_{\alpha\beta})_{\dot{A}\dot{B}}
=(J_{\alpha\beta})^{(1)}_{\dot{A}\dot{B}}
+(J_{\alpha\beta})^{(2)}_{\dot{A}\dot{B}}
\end{align}
where
\begin{align}
\label{jmtx0}
(J_{\alpha\beta})_{\dot{A}\dot{B}}^{(1)}
&=i\Psi_{\alpha a\dot{A}}\overline{\Psi}^{\beta a\dot{B}},&
(J_{\alpha\beta})_{\dot{A}\dot{B}}^{(2)}
&=i\overline{\Psi}^{\alpha a\dot{A}}
\Psi_{\beta a\dot{B}}.
\end{align}
Notice that the R-symmetry generators satisfy the relations
\begin{align}
\label{bmtx1}
(J_{++})_{\dot{A}\dot{B}}&=-(J_{++})_{\dot{B}\dot{A}}\\
\label{bmtx2}
(J_{--})_{\dot{A}\dot{B}}&=-(J_{--})_{\dot{B}\dot{A}}\\
\label{bmtx3}
(J_{+-})_{\dot{A}\dot{B}}&=-(J_{-+})_{\dot{B}\dot{A}}
\end{align}
and therefore the matrices $J_{++}$, $J_{--}$ and $J_{-+}$ contain 
$28$, $28$ and $64$ independent entries respectively 
while $J_{-+}$ yields no independent ones 
because of the relations (\ref{bmtx3}). 
Therefore the R-symmetry matrix totally carries 
$28+28+64=120$ elements.

Using the canonical (anti)commutation relations (\ref{canonical0}), 
one can find the complete set of (anti)commutators among the generators
\begin{align}
\label{osp01a}
\begin{array}{ccc}
[H,D]=iH,
&[K,D]=-iK,
&[H,K]=2iD
\cr
\end{array}
\end{align}
\begin{align}
\label{osp01b}
\begin{array}{ccc}
[(J_{\alpha\beta})_{\dot{A}\dot{B}},H]=0,
&[(J_{\alpha\beta})_{\dot{A}\dot{B}},D]=0,
&[(J_{\alpha\beta})_{\dot{A}\dot{B}},K]=0
\end{array}
\end{align}
\begin{align}
\label{osp01c}
[(J_{\alpha\beta})_{\dot{A}\dot{B}},
(J_{\gamma\delta})_{\dot{C}\dot{D}}]
&=i(J_{\gamma\beta})_{\dot{C}\dot{B}}
\delta_{\alpha\delta}\delta_{\dot{A}\dot{D}}
-i(J_{\alpha\delta})_{\dot{A}\dot{D}}
\delta_{\beta\gamma}\delta_{\dot{B}\dot{C}}
\nonumber\\
&+i(J_{\delta\beta})_{\dot{D}\dot{B}}
\delta_{\alpha\gamma}\delta_{\dot{A}\dot{C}}
-i(J_{\alpha\gamma})_{\dot{A}\dot{C}}
\delta_{\beta\delta}\delta_{\dot{B}\dot{D}}
\end{align}
\begin{align}
\label{osp01d}
\begin{array}{ccc}
[H,Q_{\alpha\dot{A}}]=0,
&[D,Q_{\alpha\dot{A}}]=-\frac{i}{2} Q_{\alpha\dot{A}},
&[K,Q_{\alpha\dot{A}}]=iS_{\alpha\dot{A}}
\cr
[H,\overline{Q}^{\alpha\dot{A}}]=0,
&[D,\overline{Q}^{\alpha\dot{A}}]
=-\frac{i}{2}\overline{Q}^{\alpha\dot{A}},
&[K,\overline{Q}^{\alpha\dot{A}}]
=i\overline{S}^{\alpha\dot{A}}
\end{array}
\end{align}
\begin{align}
\label{osp01e}
\begin{array}{ccc}
[H,S_{\alpha\dot{A}}]=-iQ_{\alpha\dot{A}},
&[D,S_{\alpha\dot{A}}]=\frac{i}{2}S_{\alpha\dot{A}},
&[K,S_{\alpha\dot{A}}]=0
\cr
[H,\overline{S}^{\alpha\dot{A}}]
=-i\overline{Q}^{\alpha\dot{A}}
&[D,\overline{S}^{\alpha\dot{A}}]
=\frac{i}{2}\overline{S}^{\alpha\dot{A}},
&[K,\overline{S}^{\alpha\dot{A}}]=0
\cr
\end{array}
\end{align}
\begin{align}
\label{osp01f}
\{Q_{\alpha\dot{A}},\overline{Q}^{\beta\dot{B}}\}
&=2H\delta_{\alpha\beta}\delta_{\dot{A}\dot{B}}\nonumber\\
\{S_{\alpha\dot{A}},\overline{S}^{\beta\dot{B}}\}
&=2K\delta_{\alpha\beta}\delta_{\dot{A}\dot{B}}\nonumber\\
\{Q_{\alpha\dot{A}},
\overline{S}^{\beta\dot{B}}\}
&=-2D\delta_{\alpha\beta}\delta_{\dot{A}\dot{B}}
+(J_{\alpha\beta})^{(1)}_{\dot{A}\dot{B}}
\left(
\delta_{\alpha\beta}
-\delta_{\alpha -\beta}
\right)
-\frac{i}{2}\delta_{\alpha\beta}\delta_{\dot{A}\dot{B}}
\nonumber\\
\{\overline{Q}^{\alpha\dot{A}},S_{\beta\dot{B}}\}
&=-2D\delta_{\alpha\beta}\delta_{\dot{A}\dot{B}}
+(J_{\alpha\beta})^{(2)}_{\dot{A}\dot{B}}
\left(
\delta_{\alpha\beta}
-\delta_{\alpha -\beta}
\right)
-\frac{i}{2}\delta_{\alpha\beta}\delta_{\dot{A}\dot{B}}
\end{align}
\begin{align}
\label{osp01g}
[(J_{\alpha\beta})_{\dot{A}\dot{B}},Q_{\gamma\dot{C}}]
&=i\left(
Q_{\beta\dot{B}}\delta_{\alpha\gamma}\delta_{\dot{A}\dot{C}}
-Q_{\alpha\dot{A}}\delta_{\beta\gamma}\delta_{\dot{B}\dot{C}}
\right)
\nonumber\\
[(J_{\alpha\beta})_{\dot{A}\dot{B}},S_{\gamma\dot{C}}]
&=i\left(
S_{\beta\dot{B}}\delta_{\alpha\gamma}\delta_{\dot{A}\dot{C}}
-S_{\alpha\dot{A}}\delta_{\beta\gamma}\delta_{\dot{B}\dot{C}}
\right)\nonumber\\
[(J_{\alpha\beta})_{\dot{A}\dot{B}},\overline{Q}^{\gamma\dot{C}}]
&=-i\left(
\overline{Q}^{\beta\dot{B}}\delta_{\alpha\gamma}\delta_{\dot{A}\dot{C}}
-\overline{Q}^{\alpha\dot{A}}\delta_{\beta\gamma}\delta_{\dot{B}\dot{C}}
\right)
\nonumber\\
[(J_{\alpha\beta})_{\dot{A}\dot{B}},\overline{S}^{\gamma\dot{C}}]
&=-i\left(
\overline{S}^{\beta\dot{B}}\delta_{\alpha\gamma}\delta_{\dot{A}\dot{C}}
-\overline{S}^{\alpha\dot{A}}\delta_{\beta\gamma}\delta_{\dot{B}\dot{C}}
\right).
\end{align}

The Hamiltonian $H$, the dilatation generator $D$ and 
the conformal boost generator $K$ satisfy 
the one-dimensional conformal algebra
(\ref{osp01a}). 
By defining
\begin{align}
\label{so12a}
T_{0}&=\frac12\left(
\frac{K}{a}+aH
\right)\\
T_{1}&=D\\
T_{2}&=\frac12\left(
\frac{K}{a}-aH
\right)
\end{align}
with $a$ being a constant with dimension of length, 
one finds the explicit representation 
of the $\mathfrak{so}(1,2)$ algebra
\begin{align}
\label{so12b}
[T_{i},T_{j}]=i\epsilon_{ijk}T^{k}
\end{align}
where $\epsilon_{ijk}$ is a three-index anti-symmetric tensor with 
$\epsilon_{012}=1$ and 
$g_{ij}=\mathrm{diag}(1,-1,-1)$. 
Alternatively if we introduce
\begin{align}
\label{vira1}
L_{0}&=\frac12 (aH+\frac{K}{a})\\
L_{\pm}&=\frac12\left(
aH-\frac{K}{a}\pm 2iD
\right),
\end{align}
then we get the explicit representation 
of the $\mathfrak{sl}(2,\mathbb{R})$ algebra in the Virasoro form
\begin{align}
\label{vira2}
[L_{m},L_{n}]=(m-n)L_{m+n}
\end{align}
with $m,n=0,\pm1$. 

As the superpartners of the conformal generators 
there are sixteen supercharges 
$Q_{\alpha\dot{A}}$ and as many 
superconformal generators $S_{\alpha\dot{A}}$. 
As seen from (\ref{osp01b}) and (\ref{osp01g}), 
the R-symmetry generators $(J_{\alpha\beta})_{\dot{A}\dot{B}}$ 
commute with the bosonic generators 
$H$, $D$ and $K$ while they yield the rotations 
of the fermionic generators $Q_{\alpha\dot{A}}$ 
and $S_{\alpha\dot{A}}$. 
The commutation relation (\ref{osp01c}) implies that 
$(J_{\alpha\beta})_{\dot{A}\dot{B}}$ 
obey the $\mathfrak{so}(16)$ algebra. 
Therefore we can conclude that 
the theory (\ref{effk1}) is the $OSp(16|2)$ invariant 
$\mathcal{N}=16$ superconformal mechanics. 
Indeed this fits in the list 
of the possible simple supergroup for 
superconformal quantum mechanics 
\cite{Claus:1998us,BrittoPacumio:1999ax}.

We see that the R-symmetry is now enhanced in our quantum mechanics. 
Interestingly a similar phenomenon has been already observed in 
$d=11$ supergravity. 
In $d=11$ supergravity 
the original tangent space symmetry $SO(1,10)$ 
can break down into the subgroup $SO(1,2)\times SO(8)$ 
through a partial choice of gauge for the elfbein. 
However, it has been pointed out in 
\cite{deWit:1985iy,Nicolai:1986jk,deWit:1986mz} that 
one can find the enhanced $SO(1,2)\times SO(16)$ tangent space symmetry 
by introducing new gauge degrees of freedom. 
It would be intriguing to inquire  
whether the enlarged R-symmetry of our quantum mechanics 
reflects that of $d=11$ supergravity.

\subsection{$\mathcal{N}=12$ superconformal mechanics}
\label{abjmscqm}
\subsubsection{Derivation of quantum mechanics}
Let us consider the $U(N)_{k}\times \hat{U}(N)_{-k}$ ABJM-model on
$\mathbb{R}\times T^{2}$. 
The theory may describe the dynamics of $N$ coincident M2-branes 
with the world-volume $M_{3}=\mathbb{R}\times T^{2}$ 
moving in a transverse space with an $SU(4)$ holonomy.  
We now want to derive the low-energy effective theory describing 
the dynamics around static BPS configurations. 
Such BPS configurations obey the BPS equations. 
From the supersymmetry transformations 
(\ref{abjms03}), (\ref{abjms04}) for fermions 
we find the following set of BPS equations: 
\begin{align}
\label{bpsabjm01}
&D_{z}Y^{A}=0,\ \ \ D_{\overline{z}}Y^{A}=0\\
\label{bpsabjm02}
&Y^{C}Y_{C}^{\dag}Y^{B}-Y^{B}Y_{C}^{\dag}Y^{C}=0\\
\label{bpsabjm03}
&Y^{C}Y_{A}^{\dag}Y^{D}=0.
\end{align}
To satisfy the algebraic equations 
(\ref{bpsabjm02}) and (\ref{bpsabjm03}), 
the bosonic Higgs fields $Y^{A}$ and $Y_{A}^{\dag}$ should take the
diagonal form
\begin{align}
\label{ydiag1}
Y^{A}&=\mathrm{diag}(y_{1}^{A}, \cdots, y_{N}^{A}),& 
Y^{\dag}_{A}&=\mathrm{diag}(\overline{y}_{A1}, \cdots,
 \overline{y}_{AN}) 
\end{align}
where $y_{a}^{A}$ is a complex scalar field. 
For the above diagonal configurations, 
all the off-diagonal elements are massive and 
the gauge group $U(N)\times \hat{U}(N)$ is spontaneously broken to 
$U(1)^{N}$ \cite{Aharony:2008ug}. 
Let us define 
\begin{align}
\label{qgauge1}
\mathcal{A}^{+}_{\mu a}&:=A_{\mu aa}+\hat{A}_{\mu aa},&
\mathcal{A}^{-}_{\mu a}&:=A_{\mu aa}-\hat{A}_{\mu aa}
\end{align}
where the indices $a=1, \cdots, N$ characterize the gauge degrees of
freedom, i.e. the internal degrees of freedom of the multiple M2-branes.  
Note that all the couplings involve the gauge fields $\mathcal{A}_{\mu a}^{-}$ 
while the other gauge fields $\mathcal{A}_{\mu a}^{+}$ are associated with 
the preserved $U(1)$ gauge group. 
In terms of the
expressions (\ref{ydiag1}) and (\ref{qgauge1}), 
we can rewrite the equations (\ref{bpsabjm01}) as 
\begin{align}
\label{bpsabjm011}
\partial_{z}y_{a}^{A}+i\mathcal{A}_{za}^{-}y_{a}^{A}&=0,&
\partial_{z}\overline{y}_{Aa}-i\mathcal{A}_{za}^{-}\overline{y}_{Aa}&=0
\\
\label{bpsabjm012}
\partial_{\overline{z}}y_{a}^{A}+i\mathcal{A}_{\overline{z}a}^{-}y_{a}^{A}&=0,& 
\partial_{\overline{z}}\overline{y}_{Aa}-i\mathcal{A}_{\overline{z}a}^{-}\overline{y}_{Aa}&=0\\
\label{bpsabjm013}
A_{z ab}=\hat{A}_{z ab}
=A_{\overline{z}ab}&=\hat{A}_{\overline{z}ab}=0
&\mathrm{for\ }\ a\neq b &.
\end{align}
The first and second lines correspond to the equations for diagonal elements 
and last one is for the off-diagonal elements. 
The general solutions to 
the equations (\ref{bpsabjm011}) and (\ref{bpsabjm012}) are given by
\begin{align}
\label{bpsabjm14}
y_{a}^{A}&=r_{a}^{A}e^{i
\left(
\varphi_{a}(z,\overline{z})+\theta^{A}_{a}
\right)}
\\
\label{bpsabjm15}
\mathcal{A}_{za}^{-}&=-\partial_{z}\varphi_{a}(z,\overline{z})
\end{align}
where $r_{a}^{A}$, $\theta_{a}^{A}$ $\in \mathbb{R}$ 
have no dependence on $z$ and $\overline{z}$ 
while $\varphi_{a}(z,\overline{z}) \in \mathbb{R}$ 
is a function of $z$ and $\overline{z}$. 
The expression (\ref{bpsabjm15}) ensures 
the flatness of the $U(1)$ gauge field $\mathcal{A}_{z}^{-}$.  
Hence $\varphi_{a}$, $\mathcal{A}_{za}^{-}$ and
$\mathcal{A}_{\overline{z}a}^{-}$ take the form 
\cite{AlvarezGaume:1986es}
\begin{align}
\label{phiabjm1}
&\varphi_{a}(z,\overline{z})
=-2\pi\frac{\Theta_{a}}{\tau-\overline{\tau}}z+
2\pi\frac{\overline{\Theta}_{a}}{\tau-\overline{\tau}}\overline{z}\\
\label{azabjm1}
&\mathcal{A}_{za}^{-}
=2\pi\frac{\Theta_{a}}{\tau-\overline{\tau}}\omega, \ \ \ \ \ \ \ 
\mathcal{A}_{\overline{z}a}^{-}
=-2\pi\frac{\overline{\Theta}_{a}}{\tau-\overline{\tau}}\overline{\omega}.
\end{align}
Here $\tau$ is the moduli of the torus 
defined in (\ref{period0}) and 
$\Theta_{a}:=\zeta_{a}+\overline{\tau}\xi_{a}$, 
$a=1, \cdots, N$ are the coordinates of the product space 
of the $N$ Jacobi varieties characterizing the $N$ $U(1)$ flat bundles. 
For the bosonic Higgs fields to 
describe the positions of the membranes, 
we should impose the single-valuedness of $y_{a}^{A}$ as
\begin{align}
\label{abjmsing1}
y_{a}^{A}(z+1,\overline{z}+1)
&=y_{a}^{A}(z,\overline{z})
\nonumber\\
y_{a}^{A}(z+\tau,\overline{z}+\overline{\tau})
&=y_{a}^{A}(z,\overline{z}).
\end{align}  
These conditions require that $\xi_{a}$ and $\zeta_{a}$ are integers, 
which result in the quantization of the variables $\varphi_{a}$,
$\mathcal{A}_{za}^{-}$ and $\mathcal{A}_{\overline{z}a}^{-}$. 
Then the resulting static BPS configurations are 
\begin{align}
\label{1bps1}
Y^{A}&=\mathrm{diag}(y_{1}^{A}, \cdots, y_{N}^{A})
=\textrm{diag}
\left(
r_{1}^{A}e^{i(\varphi_{1}(z,\overline{z})+\theta_{1}^{A})},
\cdots,
r_{N}^{A}e^{i(\varphi_{N}(z,\overline{z})+\theta_{N}^{A})}
\right)\nonumber\\
Y_{A}^{\dag}&=\mathrm{diag}(\overline{y}_{A1}, \cdots,
 \overline{y}_{AN})
=\textrm{diag}
\left(
r_{1}^{A}e^{-i(\varphi_{1}(z,\overline{z})+\theta_{1}^{A})},
\cdots,
r_{N}^{A}e^{-i(\varphi_{N}(z,\overline{z})+\theta_{N}^{A})}
\right)
\nonumber\\
A_{z}&=\textrm{diag}
\left(
A_{z 11},\cdots,A_{z NN}
\right)\nonumber\\
\hat{A}_{z}&=A_{z}+\partial_{z}\varphi
=\textrm{diag}\left(
A_{z 11}+\partial_{z}\varphi_{1},
\cdots,A_{z NN}+\partial_{z}\varphi_{N}
\right).
\end{align}
By the supersymmetry the above bosonic configurations 
are paired with the fermionic fields
\begin{align}
\label{1bps4}
\psi_{\pm A}&=
\mathrm{diag}\left(
\psi_{\pm A 1}, \cdots, \psi_{\pm A N}
\right),&
\psi^{\dag A}_{\pm}&=
\mathrm{diag}\left(
\psi_{\pm}^{\dag A 1}, \cdots, \psi_{\pm}^{\dag A N} 
\right)
\end{align}
where the subscripts $\pm$ label the $SO(2)_{E}$ spinor representation.

Inserting the set of BPS configurations 
(\ref{1bps1}) and (\ref{1bps4}) 
into the ABJM action (\ref{abjmlag1}) 
one finds
\begin{align}
\label{abjmeff1a}
S=
\int_{\mathbb{R}} dt\int_{T^{2}}d^{2}z 
\sum_{A}
\sum_{a=1}^{N}\Biggl[
&D_{0}\overline{y}_{A}^{a}D_{0}y_{a}^{A}
-i\psi_{+}^{\dag Aa}D_{0}\psi_{+Aa}
-i\psi_{-}^{\dag Aa}D_{0}\psi_{-Aa}
\nonumber\\
&
+\frac{k}{4\pi}
\left(
\mathcal{A}^{-}_{0a}\mathcal{F}_{z\overline{z}a}^{+}
+\frac12 \mathcal{A}_{\overline{z}a}^{-}\dot{\mathcal{A}}_{za}^{+}
-\frac12 \mathcal{A}_{za}^{-}\dot{\mathcal{A}}_{\overline{z}a}^{+}
\right)
\Biggr].
\end{align}
Recall that $\mathcal{A}_{z}^{-}$ and $\mathcal{A}_{\overline{z}}^{-}$ 
are quantized and their time derivative terms do not show up in the action. 
Thus we can treat them as auxiliary fields and integrate out them. 
Consequently we get constraints 
$\dot{\mathcal{A}}_{za}^{+}=\dot{\mathcal{A}}_{\overline{z}a}^{+}=0$, 
which imply that 
the gauge fields $\mathcal{A}_{za}^{+}$ and $\mathcal{A}_{\overline{z}a}^{+}$ 
on the Riemann surface have no time dependence.  

Taking these constraints into account and 
proceeding the integration over the torus, 
we obtain the low-energy effective action 
\begin{align}
\label{effss2}
S=\int_{\mathbb{R}}dt 
\left[
D_{0}\overline{y}^{a}_{A}D_{0}y_{a}^{A}
-i\psi^{\dag \alpha Aa}D_{0}\psi_{\alpha Aa}
+kC_{1}(E_{a})\mathcal{A}_{0a}^{-}
\right].
\end{align}
Here the repeated indices are summed over and 
$\alpha,\beta,\cdots=+,-$ denote the $SO(2)_{E}$ spinor indices. 
The covariant derivatives are defined by
\begin{align}
\label{abjmgcov1}
D_{0}y^{A}_{a}
&=\dot{y}_{a}^{A}+i\mathcal{A}^{-}_{0a}y_{a}^{A},&
D_{0}\overline{y}_{Aa}
&=\dot{\overline{y}}_{Aa}-i\mathcal{A}_{0a}^{-}\overline{y}_{Aa}
\nonumber\\
D_{0}\psi_{\alpha Aa}
&=\dot{\psi}_{\alpha Aa}+i\mathcal{A}^{-}_{0a}\psi_{\alpha Aa},& 
D_{0}\psi_{\alpha a}^{\dag A}&=\dot{\psi}^{\dag A}_{\alpha a}
-i\mathcal{A}^{-}_{0a}\psi^{\dag A}_{\alpha a}.
\end{align}
and
\begin{align}
\label{ch02}
C_{1}(E_{a}):=
\frac{1}{2\pi}\int_{T^{2}}
F_{z\overline{z}aa}
=\frac{1}{4\pi}\int_{T^{2}}
\mathcal{F}^{+}_{z\overline{z} a}
\end{align}
is the Chern number of the $a$-th $U(1)$ principal bundle 
$E_{a}\rightarrow T^{2}$ over the torus 
associated with 
the preserved $U(1)$ gauge fields $A_{z aa}$. 

The action (\ref{effss2}) is invariant 
under the one-dimensional conformal transformations
\begin{align}
\label{tconf2}
\delta t&=f(t)=a+bt+ct^{2}, 
&\delta \partial_{0}&=-\dot{f}\partial_{0}\\
\delta y_{a}^{A}&=\frac12 \dot{f}y_{a}^{A}, 
&\delta \overline{y}_{Aa}&=\frac12 \dot{f}\overline{y}_{Aa}\\
\delta\psi_{\alpha Aa}&=0, 
&\delta\psi^{\dag A}_{\alpha a}&=0\\
\delta \mathcal{A}_{0a}^{-}&=-\dot{f}\mathcal{A}_{0a}^{-}
\end{align}
and $\mathcal{N}=12$ supersymmetry transformations
\begin{align}
\label{tsusy03}
\delta
 y_{a}^{A}&=i\omega^{\alpha AB}\psi_{\alpha Ba},&
\delta
\overline{y}_{Aa}&=i\psi^{\dag \alpha B}_{a}\omega_{\alpha AB}\\
\delta\psi_{\alpha Aa}
&=\omega_{\alpha AB}D_{0}y_{a}^{B},&
\delta\psi_{\alpha a}^{\dag A}
&=-D_{0}\overline{y}_{Ba}\omega_{\alpha}^{AB}\\
\delta\mathcal{A}^{-}_{0a}
&=0
\end{align}
where the supersymmetry parameters 
$\omega_{+AB}:=\epsilon_{+i}(\Gamma^{i})_{AB}$ 
and $\omega_{-AB}:=\epsilon_{-i}(\Gamma^{i})_{AB}$ transform as 
$\bm{6}_{+}$ and $\bm{6}_{-}$ under $SU(4)\times SO(2)_{E}$ respectively. 
Therefore the low-energy effective theory is described by 
the $\mathcal{N}=12$ superconformal gauged quantum mechanics (\ref{effss2}).

\subsubsection{Reduced system with inverse-square interaction}
\label{abjmscqm1}
The low-energy effective action (\ref{effss2}) 
is quadratic in $\mathcal{A}_{0a}^{-}$ 
and contains no time derivatives of $\mathcal{A}_{0a}^{-}$. 
So they are auxiliary fields and we want to integrate them out. 
Let us fix the gauge as $\mathcal{A}_{0a}^{-}=0$.  
Then the algebraic equations of motion of $\mathcal{A}^{-}_{0a}$ yield 
the Gauss law constraints, the moment map conditions
\begin{align}
\label{1a1}
\phi_{0a}:=kC_{1}(E_{a})
+2\sum_{A}(r_{a}^{A})^{2}\dot{\theta}_{a}^{A}
+\sum_{A}
\psi^{\dag \alpha A a}\psi_{\alpha A a}
=0
\end{align}
for $a=1, \cdots, N$. 
Note that although the set of equations (\ref{1a1}) has the same form as 
that of (\ref{1a0}), 
the physical meaning of these constraints are different 
because the angular variable $\theta_{a}^{A}$'s  
are defined not in the abstract space of the internal degrees of freedom 
as in (\ref{1a0}), 
but in the actual configuration space of the $a$-th M2-brane 
in the $A$-th complex plane. 

Defining the conserved charges 
$h_{a}^{A}:=2(r_{a}^{A})^{2}\dot{\theta}_{a}^{A}$, 
using the above constraints (\ref{1a1}) 
and following the reduction procedure as in the derivation of 
(\ref{0eff0}), 
we can integrate out the auxiliary gauge fields $\mathcal{A}_{0a}^{-}$ 
and find the reduced effective action 
with the inverse-square type interaction
\begin{align}
\label{abjmeff1b}
S=\int_{\mathbb{R}}dt 
&\sum_{a=1}^{N}
\Biggl[
\dot{x}^{2}_{a}
-\frac{i}{2}\sum_{A\neq B}
\left(
\psi^{\dag\alpha A a}\dot{\psi}_{\alpha A a}
-\dot{\psi}^{\dag Aa}\psi_{\alpha Aa}
\right)
\nonumber\\
&
+\sum_{A\neq B}(\dot{r}_{a}^{A})^{2}
-\frac{i}{2}
\left(
\lambda^{\dag \alpha a}\dot{\lambda}_{\alpha a}
-\dot{\lambda}^{\dag\alpha a}\lambda_{\alpha a}
\right)
\nonumber\\
&
-\frac{\left[
kC_{1}(E_{a})
+\sum_{A\neq B}h_{a}^{A}
+\sum_{A\neq B}
\psi^{\dag \alpha A a}\psi_{\alpha A a}
+\lambda^{\dag \alpha a}\lambda_{\alpha a}
\right]^{2}}
{4x_{a}^{2}}
-\sum_{A\neq B}\frac{(h_{a}^{A})^{2}}{4(r_{a}^{A})^{2}}
\Biggr].
\end{align}
Here $x_{a}:=r_{a}^{B}$ describes the motion of 
the $a$-th M2-brane in the $B$-th complex plane 
in which the corresponding ``angular momentum'' $h_{a}^{B}$ is determined 
by the assignment of the other preserved charges.  
We have also introduced the fermionic variable 
$\lambda_{\alpha a}:=\psi_{\alpha B a}$ with $A=B$, 
which turns out to be the superpartner of 
$r_{a}^{C}$, $C=1,2,3$, 
as we will see the supersymmetry transformations 
(\ref{abjmqmsusy05}) and (\ref{abjmqmsusy06}).

The action (\ref{abjmeff1b}) leads to the following 
equations of motion
\begin{align}
\label{eomabjm00}
\ddot{x}_{a}
&=\frac{\left[
kC_{1}(E_{a})
+\sum_{A\neq B}h_{a}^{A}
+\sum_{A\neq B}\psi^{\dag\alpha Aa}\psi_{\alpha Aa}
+\lambda^{\dag\alpha a}\lambda_{\alpha a}
\right]^{2}}
{4x_{a}^{3}}\\
\label{eomabjm01}
\ddot{r}_{a}^{A}
&=\frac{(h_{a}^{A})^{2}}
{4(r_{a}^{A})^{3}}\\
\label{eomabjm02}
\dot{\psi}_{\alpha Aa}
&=i\frac{
kC_{1}(E_{a})
+\sum_{A\neq B}h_{a}^{A}
+\sum_{A\neq B}\psi^{\dag\alpha Aa}\psi_{\alpha Aa}
+\lambda^{\dag\alpha a}\lambda_{\alpha a}}
{2x_{a}}\psi_{\alpha A a}
\\
\label{eomabjm03}
\dot{\psi}^{\dag \alpha Aa}
&=-i
\frac{kC_{1}(E_{a})
+\sum_{A\neq B}h_{a}^{A}
+\sum_{A\neq B}\psi^{\dag\alpha Aa}\psi_{\alpha Aa}
+\lambda^{\dag\alpha a}\lambda_{\alpha a}}
{2x_{a}}\psi^{\dag\alpha Aa}
\\
\label{eomabjm04}
\dot{\lambda}_{\alpha a}
&=i\frac{
kC_{1}(E_{a})
+\sum_{A\neq B}h_{a}^{A}
+\sum_{A\neq B}\psi^{\dag\alpha Aa}\psi_{\alpha Aa}
+\lambda^{\dag\alpha a}\lambda_{\alpha a}}
{2x_{a}}\lambda_{\alpha a}\\
\label{eomabjm05}
\dot{\lambda}^{\dag \alpha a}
&=-i
\frac{kC_{1}(E_{a})
+\sum_{A\neq B}h_{a}^{A}
+\sum_{A\neq B}\psi^{\dag\alpha Aa}\psi_{\alpha Aa}
+\lambda^{\dag\alpha a}\lambda_{\alpha a}}
{2x_{a}}\lambda^{\dag\alpha a}.
\end{align}
Using the fermionic equations of motion 
(\ref{eomabjm02})-(\ref{eomabjm05}), 
we can check that the Gauss law constraint (\ref{1a1}) 
has no time dependence, i.e. $\dot{\phi}_{0a}=0$.

The canonical momenta are given by
\begin{align}
\label{mom11}
p^{a}&:=\frac{\partial L}{\partial \dot{x}_{a}}
=2\dot{x}^{a},& 
P^{a}_{A}&:=\frac{\partial L}{\partial \dot{r}_{a}^{A}}
=2\dot{r}^{a}_{A}\\
\label{mom12}
\pi^{\alpha Aa}
&:=\frac{\vec{\partial}L}{\partial \dot{\psi}_{\alpha Aa}}
=\frac{i}{2}\psi^{\dag \alpha A a},&
\tilde{\pi}_{\alpha Aa}
&:=\frac{\vec{\partial}L}{\partial \dot{\psi}^{\dag\alpha Aa}}
=\frac{i}{2}\psi_{\alpha Aa}
\\
\label{mom13}
\Pi^{\alpha a}
&:=\frac{\vec{\partial}L}{\partial \dot{\lambda}_{\alpha a}}
=\frac{i}{2}\lambda^{\dag \alpha a},&
\tilde{\Pi}_{\alpha a}
&:=\frac{\vec{\partial}L}{\partial \dot{\lambda}^{\dag \alpha a}}
=\frac{i}{2}\lambda_{\alpha a}.
\end{align}
The fermionic canonical momenta provide 
the second class constraints
\begin{align}
\label{prcons2}
\phi_{1}^{\alpha Aa}
&:=\pi^{\alpha Aa}-\frac{i}{2}\psi^{\dag Aa}=0,&
\phi_{2\alpha Aa}
&:=\tilde{\pi}_{\alpha Aa}-\frac{i}{2}\psi_{\alpha Aa}=0\\
\label{prcons3}
\phi_{3}^{\alpha a}
&:=\Pi^{\alpha a}-\frac{i}{2}\lambda^{\dag \alpha a}=0,&
\phi_{4\alpha a}
&:=\tilde{\Pi}_{\alpha a}-\frac{i}{2}\lambda_{\alpha a}=0.
\end{align}
Taking account into the constraints (\ref{prcons2}) and (\ref{prcons3}), 
we find the Dirac brackets
\begin{align}
\label{dirac3}
[x_{a},p^{b}]_{DB}&=\delta_{ab},&
[r_{a}^{A},P_{B}^{b}]_{DB}=\delta_{AB}\delta_{ab}\\
\label{dirac4}
\left[
\psi_{\alpha Aa},\psi^{\dag \beta Bb}
\right]_{DB}
&=i\delta_{\alpha\beta}\delta_{AB}\delta_{ab},&
\left[
\lambda_{\alpha a},\lambda^{\dag\beta b}
\right]_{DB}
=i\delta_{\alpha\beta}\delta_{ab}.
\end{align}

The action (\ref{abjmeff1b}) possesses 
the one-dimensional conformal invariance 
\begin{align}
\label{abjmconf00}
\delta t&=f(t)=a+bt+ct^{2}, 
&\delta \partial_{0}&=-\dot{f}\partial_{0}\\
\delta x_{a}&=\frac12 \dot{f}x_{a}, 
&\delta r_{a}^{A}&=\frac12 \dot{f}r_{a}^{A}\\
\delta\psi_{\alpha Aa}&=0, 
&\delta\psi^{\dag \alpha A}_{a}&=0\\
\delta\lambda_{\alpha a}&=0, 
&\delta\lambda^{\dag\alpha}_{a}&=0.
\end{align}
Using the Noether's procedure 
we find the $SL(2,\mathbb{R})$ generators
\begin{align}
\label{abjmconf01}
H&=\sum_{a=1}^{N}
\Biggl[
\frac{p_{a}^{2}}{4}
+\frac{\left(
kC_{1}(E_{a})
+\sum_{A\neq B}h_{a}^{A}
+\sum_{A}
\psi^{\dag \alpha A a}\psi_{\alpha A a}
+\lambda^{\dag \alpha a}\lambda_{\alpha a}
\right)
^{2}}{4x_{a}^{2}}\nonumber\\
&+\sum_{A\neq B}\left(
\frac{(P_{a}^{A})^{2}}{4}
+\frac{(h_{a}^{A})^{2}}{4(r_{a}^{A})^{2}}
\right)
\Biggr]\\
\label{abjmconf02}
D&=-\frac14 \sum_{a=1}^{N}
\left[
\left\{x_{a},p_{a}\right\}
+\sum_{A\neq B}
\left\{r_{a}^{A},P_{a}^{A}\right\}
\right]\\
\label{abjmconf03}
K&=\sum_{a=1}^{N}\left[
x_{a}^{2}+\sum_{A\neq B}(r_{a}^{A})^{2}
\right].
\end{align}
Here we have absorbed the time dependent part 
of $D$ and $K$ by 
similarity transformations (\ref{simdk0}).  

Also the action (\ref{abjmeff1b}) is invariant 
under the following fermionic transformations 
\begin{align}
\label{abjmqmsusy01}
\delta x_{a}&=
\frac{i}{\sqrt{2}}
\left(
\epsilon^{\alpha C}\psi_{\alpha Ca}
+\epsilon^{\dag}_{\alpha C}\psi^{\dag\alpha C}_{a}
\right)
\\
\label{abjmqmsusy02}
\delta r_{a}^{C}&=
\frac{i}{2}
\left[
\left(
\omega^{\alpha CD}\psi_{\alpha Da}
\right)
e^{-i\theta_{a}^{C}}
+
\left(
\psi^{\dag \alpha D}_{a}\omega_{\alpha CD}
\right)e^{i\theta_{a}^{C}}
-
\left(\epsilon^{\alpha C}\lambda_{\alpha a}
\right)e^{-i\theta_{a}^{C}}
-
\left(\epsilon_{\alpha C}^{\dag}\lambda^{\dag\alpha}_{a}
\right)e^{i\theta_{a}^{C}}
\right]
\\
\label{abjmqmsusy03}
\delta \psi_{\alpha Ca}&=
\left(
\dot{r}_{a}^{D}+i\frac{h_{a}^{D}}{2r_{a}^{D}}
\right)e^{i\theta_{a}^{D}}\omega_{\alpha CD}\nonumber\\
&+
\sqrt{2}\left(
\dot{x}_{a}
-i\frac{kC_{1}(E_{a})
+\sum_{D\neq B}h_{a}^{D}
+\psi^{\dag \alpha Da}\psi_{\alpha Da}
+\lambda^{\dag \alpha a}\lambda_{\alpha a}}{2x_{a}}
\right)\epsilon_{\alpha C}^{\dag}
-\frac{i}{\sqrt{2}}
\frac{l_{a}}{x_{a}}\psi_{\alpha Ca}
\\
\label{abjmqmsusy04}
\delta \psi^{\dag\alpha C}_{a}&=
-\left(
\dot{r}_{a}^{D}-i\frac{h_{a}^{D}}{2r_{a}^{D}}
\right)e^{-i\theta_{a}^{D}}\omega_{\alpha}^{CD}\nonumber\\
&+
\sqrt{2}\left(
\dot{x}_{a}
+i\frac{
kC_{1}(E_{a})
+\sum_{D\neq B}h_{a}^{D}
+\psi^{\dag \alpha Da}\psi_{\alpha Da}
+\lambda^{\dag\alpha a}\lambda_{\alpha a}
}
{2x_{a}}
\right)\epsilon^{\alpha C}
+\frac{i}{\sqrt{2}}
\frac{l_{a}}{x_{a}}\psi_{a}^{\dag\alpha C}
\\
\label{abjmqmsusy05}
\delta \lambda_{\alpha a}
&=
-\epsilon_{\alpha C}^{\dag}
\left(\dot{r}_{a}^{C}+i\frac{h_{a}^{C}}
{2r_{a}^{C}}
\right)e^{i\theta^{C}_{a}}\\
\label{abjmqmsusy06}
\delta \lambda^{\dag\alpha}_{a}
&=-\left(
\dot{r}_{a}^{C}-i\frac{h_{a}^{C}}
{2r_{a}^{C}}
\right)e^{-i\theta_{a}^{C}}\epsilon^{\alpha C}
\end{align}
with $C,D=1,2,3$. 
Here $\epsilon^{\alpha C}$ and 
their Hermitian conjugate $\epsilon_{\alpha C}^{\dag}$ 
are infinitesimal fermionic parameters and we have defined 
\begin{align}
\label{abjmpa1}
\theta_{a}^{C}(t)&=h^{C}_{a}\int^{t} \frac{dt'}{(r^{C}_{a}(t'))^{2}}\\
\label{abjmpa2}
l_{a}&=\epsilon\psi_{a}-\epsilon^{\dag}\psi_{a}^{\dag}.
\end{align}

\subsubsection{$SU(1,1|6)$ superconformal mechanics}
\label{abjmscqm2}
The fact that the transformations
(\ref{abjmqmsusy01})-(\ref{abjmqmsusy06}) 
involve the non-local quantities 
suggests that there may exist infinitely many conserved non-local charges. 
However, we see from (\ref{abjmconf01}) 
that the Hamiltonian describing 
the motion in the $B$-th complex plane associated with 
the variable $x_{a}$ and the local charges 
commute with the others associated with the variables $r_{a}^{C}$'s 
and the non-local charges. 
Therefore they are decoupled with one another 
and we thus can analyze the dynamics in the $B$-th direction
separately. 
As in the subsection \ref{secflat3}, 
it is convenient to assign the conserved charges $h_{a}^{A}$ and 
$\lambda^{\dag\alpha a}\lambda_{\alpha a}$ to be zeros. 
Then the low-energy dynamics 
in the $B$-th complex plane is described by the action 
\begin{align}
\label{abjmscqm0}
S&=\int_{\mathbb{R}}dt \sum_{a=1}^{N}
\Biggl[
\dot{x}_{a}^{2}-i\psi^{\dag\alpha Aa}\dot{\psi}_{\alpha Aa}
-\frac{\left(
kC_{1}(E_{a})+\psi^{\dag\alpha Aa}\psi_{\alpha Aa}
\right)^{2}}
{4x_{a}^{2}}
\Biggr]
\end{align}
where $A=1,2,3$ denote the R-symmetry indices. 
Note that the action (\ref{abjmscqm0}) 
has the same structure argued in 
\cite{Ivanov:1988it,Wyllard:1999tm,Fedoruk:2011aa} 
for $\mathcal{N}>4$ superconformal quantum mechanics. 
 
The action (\ref{abjmscqm0}) has the invariance under 
the $\mathcal{N}=12$ supersymmetry transformation laws
\begin{align}
\label{su116susy1}
\delta x_{a}&=
\frac{i}{\sqrt{2}}
\left(
\epsilon^{\alpha A}\psi_{Aa}^{\alpha}
+\epsilon^{\dag}_{\alpha A}\psi_{a}^{\dag\alpha A}
\right)
\\
\label{su116susy2}
\delta\psi_{\alpha Aa}
&=
\sqrt{2}
\left(
\dot{x}_{a}-i\frac{g_{a}}{2x_{a}}
\right)\epsilon_{\alpha A}^{\dag}
-\frac{i}{\sqrt{2}}\frac{l_{a}}{x_{a}}\psi_{\alpha Aa}
\\
\label{su116susy3}
\delta\psi^{\dag \alpha A}_{a}
&=
\sqrt{2}
\left(
\dot{x}_{a}+i\frac{g_{a}}{2x_{a}}
\right)\epsilon^{\alpha A}
+\frac{i}{\sqrt{2}}\frac{l_{a}}{x_{a}}
\psi_{a}^{\dag \alpha A}
\end{align}
where 
\begin{align}
\label{g02}
g_{a}=kC_{1}(E_{a})
+\psi^{\dag \alpha Aa}\psi_{\alpha Aa}.
\end{align}
The supersymmetry transformations 
(\ref{su116susy1})-(\ref{su116susy3}) 
are generated by the supercharges
\begin{align}
\label{tsch2}
Q_{\alpha A}&=
\frac{i}{\sqrt{2}}
\left(
p^{a}-\frac{g_{a}}{x_{a}}
\right)
\psi_{\alpha Aa}\\
\tilde{Q}^{\alpha A}&=
\frac{i}{\sqrt{2}}
\left(
p^{a}+\frac{g_{a}}{x_{a}}
\right)
\psi^{\dag \alpha A}.
\end{align}

Also the action (\ref{abjmscqm0}) 
has the one-dimensional conformal invariance. 
The corresponding Noether charges are now expressed as
\begin{align}
\label{hkdconf11}
H&=\sum_{a=1}^{N}
\left[
\frac{p_{a}^{2}}{4}
+\frac{
\left(
kC_{1}(E_{a})
+\psi^{\dag \alpha Aa}\psi_{\alpha A a}
\right)^{2}}
{4x_{a}^{2}}
\right]\\
\label{hkdcon12}
D&=-\frac14\sum_{a=1}^{N}\left\{x_{a},p^{a}\right\}\\
\label{hkdcon13}
K&=\sum_{a=1}^{N}x_{a}^{2}.
\end{align}

According to the Dirac brackets (\ref{dirac3}) and (\ref{dirac4}), 
quantum operators of the canonical coordinates 
and momenta obey the quantum brackets
\begin{align}
\label{canonical1}
[x_{a},p^{b}]&=i\delta_{ab},&
\left\{
\psi_{\alpha Aa},\psi^{\dag \beta Bb}
\right\}=-\delta_{\alpha\beta}\delta_{AB}\delta_{ab}.
\end{align}

Combining the supercharges and 
the conformal generators, 
we find the superconformal boost generators
\begin{align}
\label{scboost2}
S_{\alpha A}&=
\sqrt{2}i\sum_{a}x_{a}\psi_{\alpha Aa}\\
\tilde{S}^{\alpha A}
&=\sqrt{2}i\sum_{a}x_{a}\psi^{\dag \alpha A}_{a}.
\end{align}
The R-symmetry generator is given by
\begin{align}
\label{j2}
(J_{\alpha\beta})_{AB}
=i\sum_{a}\psi^{\dag\beta B}_{a}\psi_{\alpha A a}.
\end{align}
Note that (\ref{j2}) is a complex $6\times 6$ matrix 
with $\alpha,\beta=+,-$ and 
$A,B=1,2,3$ and it contains $36$ complex valued elements.

Under the canonical relations (\ref{canonical1}), 
the generators form the following algebra
\begin{align}
\label{su116a}
\begin{array}{ccc}
[H,D]=iH,
&[K,D]=-iK,
&[H,K]=2iD
\cr
\end{array}
\end{align}
\begin{align}
\label{su116b}
\begin{array}{ccc}
[(J_{\alpha\beta})_{AB},H]=0,
&[(J_{\alpha\beta})_{AB},D]=0,
&[(J_{\alpha\beta})_{AB},K]=0
\end{array}
\end{align}
\begin{align}
\label{su116c}
[(J_{\alpha\beta})_{AB},
(J_{\gamma\delta})_{CD}]
&=i(J_{\alpha\delta})_{AD}
\delta_{\beta\gamma}\delta_{BC}
-i(J_{\gamma\beta})_{CB}
\delta_{\alpha\delta}\delta_{AD}
\end{align}
\begin{align}
\label{su116d}
\begin{array}{ccc}
[H,Q_{\alpha A}]=0,
&[D,Q_{\alpha A}]=-\frac{i}{2} Q_{\alpha A},
&[K,Q_{\alpha A}]=iS_{\alpha A}
\cr
[H,\tilde{Q}^{\alpha A}]=0,
&[D,\tilde{Q}^{\alpha A}]
=-\frac{i}{2}\tilde{Q}^{\alpha A},
&[K,\tilde{Q}^{\alpha A}]
=i\tilde{S}^{\alpha A}
\end{array}
\end{align}
\begin{align}
\label{su1161e}
\begin{array}{ccc}
[H,S_{\alpha A}]=-iQ_{\alpha A},
&[D,S_{\alpha A}]=\frac{i}{2}S_{\alpha A},
&[K,S_{\alpha A}]=0
\cr
[H,\tilde{S}^{\alpha A}]
=-i\tilde{Q}^{\alpha A}
&[D,\tilde{S}^{\alpha A}]
=\frac{i}{2}\tilde{S}^{\alpha A},
&[K,\tilde{S}^{\alpha A}]=0
\cr
\end{array}
\end{align}
\begin{align}
\label{su116f}
\{Q_{\alpha A},\tilde{Q}^{\beta B}\}
&=2H\delta_{\alpha\beta}\delta_{AB}\nonumber\\
\{S_{\alpha A},\tilde{S}^{\beta B}\}
&=2K\delta_{\alpha\beta}\delta_{AB}\nonumber\\
\{Q_{\alpha A},\tilde{S}^{\beta B}\}
&=-2D\delta_{\alpha\beta}\delta_{AB}
-2(J_{\alpha\beta})_{AB}
+\frac{i}{2}(2\sum_{a}g_{a}+1)\delta_{\alpha\beta}\delta_{AB}
\nonumber\\
\{\tilde{Q}^{\alpha A},S_{\beta B}\}
&=-2D\delta_{\alpha\beta}\delta_{AB}
-2(J_{\alpha\beta}^{\dag})_{AB}
-\frac{i}{2}(2\sum_{a}g_{a}+1)\delta_{\alpha\beta}\delta_{AB}
\end{align}
\begin{align}
\label{su116g}
[(J_{\alpha\beta})_{AB},Q_{\gamma C}]
&=iQ_{\alpha A}\delta_{\beta\gamma}\delta_{BC},&
[(J_{\alpha\beta})_{AB},S_{\gamma,C}]
&=iS_{\alpha A}\delta_{\beta\gamma}\delta_{BC}
\nonumber\\
[(J_{\alpha\beta})_{AB},\tilde{Q}^{\gamma C}]
&=-i\tilde{Q}^{\alpha A}\delta_{\beta\gamma}\delta_{BC},&
[(J_{\alpha\beta})_{AB},\tilde{S}^{\gamma,C}]
&=-i\tilde{S}^{\alpha A}\delta_{\beta\gamma}\delta_{BC}.
\end{align}

The Hamiltonian $H$, 
the dilatation generator $D$ and 
the conformal boost generator form the 
one-dimensional conformal algebra 
$\mathfrak{so}(1,2)=\mathfrak{sl}(2,\mathbb{R})=\mathfrak{su}(1,1)$. 
As each of the supercharges $Q_{\alpha A}$ 
and $\tilde{Q}^{\alpha A}=-(Q_{\alpha A})^{\dag}$ 
contain six real components, 
there exist twelve supercharges. 
They are the square roots of the Hamiltonian $H$. 
In addition, there are as many superconformal charges 
$S_{\alpha A}$ and $\tilde{S}^{\alpha A}$, 
which are the square roots of the conformal boost generator $K$. 
The anti-commutators of the fermionic charges generate 
an extra bosonic R-symmetry generators $(J_{\alpha\beta})_{AB}$. 
They form the $\mathfrak{u}(6)$ algebra (\ref{su116c}). 
Thus the action (\ref{abjmscqm0}) describes 
the $SU(1,1|6)$ invariant $\mathcal{N}=12$ superconformal mechanics.  
In fact this belongs to the list of the simple supergroup 
for superconformal quantum mechanics 
\cite{Claus:1998us,BrittoPacumio:1999ax}.

\section{Curved M2-branes and topological twisting}
\label{seccurv}
\subsection{M2-branes wrapping a holomorphic curve}
\label{seccurv1}
The BLG action (\ref{blglagrangian}) 
and the ABJM action (\ref{abjmlag1}) may 
describe the dynamics 
of probe membranes propagating in 
a fixed background geometry with an $SO(8)$ 
and an $SU(4)$ holonomy respectively.  
For both cases, 
the world-volume $M_{3}$ is considered as 
a flat space-time $\mathbb{R}^{1,2}$ or $\mathbb{R}\times T^{2}$.   
Now let us consider more general situations 
where curved M2-branes reside in some fixed curved background geometries. 
If we naively put the theory on a general three dimensional manifold, 
all supersymmetries are broken. 
However, 
here we shall wrap the M2-branes 
on a Riemann surface $\Sigma_{g}$ of genus $g$ 
that preserves supersymmetry 
(i.e. supersymmetric two-cycles) as the form
\begin{equation}
\label{m2geo}
 M_{3}=\mathbb{R}\times (\Sigma_{g}\subset X)
\end{equation}
where $\mathbb{R}$ is viewed as a time direction 
and $X$ is a real $2(n+1)$-dimensional space preserving supersymmetry 
with vanishing three-form gauge field. 
The only known supersymmetric two-cycles, i.e. calibrated two-cycles 
in special holonomy backgrounds 
are holomorphic curves in Calabi-Yau spaces and the corresponding
two-form calibrations are K\"{a}hler calibrations. 
So we take the ambient space $X$ as an $(n+1)$-dimensional 
Calabi-Yau space and the other space as flat. 
The geometry of the M-theory is of the form
\begin{align} 
\label{m2geo1}
\mathbb{R}^{1,8-2n}\times  CY_{n+1}.
\end{align}

\subsubsection{Supersymmetry in Calabi-Yau space}
\label{seccurv1a}
In order to count the number of preserved supersymmetries in our setup, 
we firstly need to know the dimension of the vector space 
formed by the corresponding Killing spinor $\epsilon$, 
that is the amount of supersymmetries in the background geometry. 
Since we are now considering the background geometries 
with vanishing four-form flux, 
the Killing spinor equation is given by
\begin{align}
\label{killspinor1}
\nabla_{M}\epsilon=
\left(\partial_{M}+\frac14 \omega_{MPQ}\Gamma^{PQ}\right)\epsilon=0
\end{align}
where $\omega_{MPQ}$, 
$M,N,P,Q=0,1,\cdots,10$ is an eleven-dimensional 
Levi-Civita spin connection. 
This leads to the integrability condition
\begin{align}
\label{killspinor2}
[\nabla_{M},\nabla_{N}]\epsilon
=\frac14 R_{MNPQ}\Gamma^{PQ}\epsilon=0. 
\end{align}
The equation (\ref{killspinor2}) implies that 
a Killing spinor $\epsilon$ transforms as a singlet under the 
restricted holonomy group $H\subset Spin(1,10)$ 
generated by $R_{MNPQ}\Gamma^{PQ}$. 
In other words, 
the amount of preserved supersymmetries in the special holonomy manifold 
is equivalent to the number of singlets 
in the decomposition 
of the spinor representation $\bm{32}$ of $Spin(1,10)$ into 
the representation of the holonomy group $H$. 
In our case the background geometries are taken as 
Calabi-Yau $(n+1)$-folds with the holonomy $H=SU(n+1)$, $n=1,2,3,4$ 
and the decompositions are as follows.
\begin{enumerate}
\item $CY_{5}$

In this case the geometry is of the form $\mathbb{R}\times CY_{5}$. 
This splits the $Spin(10)$ into $SU(5)$ and 
the corresponding decomposition of the spinor representation is given by
\begin{align}
\label{cy5susy}
\bm{16}&=\bm{10}_{-}\oplus \overline{\bm{5}}_{3}\oplus \bm{1}_{-5}
\nonumber\\
\bm{16}'&=\bm{10}_{+}\oplus \bm{5}_{-3}\oplus \bm{1}_{5}.
\end{align}
The existence of two singlets implies that 
the space $\mathbb{R}\times CY_{5}$ preserves 
$\frac{2}{32}=\frac{1}{16}$ supersymmetries. 

Let us define an explicit set of projections 
defining the Killing spinors. 
To this end we need to specify how 
the Calabi-Yau spaces live in the 
eleven-dimensional space-time. 
We shall consider the situations 
where the Calabi-Yau manifolds fill in the order 
$(x^{1},x^{2})$, $(x^{9},x^{10})$, $(x^{7},x^{8})$, 
$(x^{5},x^{6})$ and $(x^{3},x^{4})$. 
Then the Killig spinors can be defined 
by the eigenvalues $\pm 1$ 
for the following set of commuting matrices 
\begin{align}
\label{projcy}
\Gamma^{12910},\ \ \ \ \ 
\Gamma^{91078},\ \ \ \ \ 
\Gamma^{7856},\ \ \ \ \ 
\Gamma^{5634}.
\end{align}
The corresponding Killing spinors for $CY_{5}$ 
can be defined by the projection
\begin{align}
\label{cy5susy2}
\Gamma^{12910}\epsilon=\Gamma^{91078}\epsilon
=\Gamma^{7856}\epsilon=\Gamma^{5634}\epsilon=-\epsilon.
\end{align}
Note that this implies that $\Gamma^{012}\epsilon=\epsilon$. 
\item $CY_{4}$

For this case the geometry is 
the product form $\mathbb{R}^{1,2}\times CY_{4}$. 
This leads to the decomposition of the 
$Spin(8)$ into $SU(4)$ and that of the spinor representation 
\begin{align}
\label{cy4susy}
\bm{8}_{s}&=\bm{6}_{0}\oplus \bm{1}_{2}\oplus \bm{1}_{-2}\nonumber\\
\bm{8}_{c}&=\bm{4}_{-}\oplus \bm{4}_{+}.
\end{align}
We see that the decomposition provides 
two singlets from sixteen components. 
Thus the geometry $\mathbb{R}^{1,2}\times CY_{4}$ 
can preserve $\frac{2}{16}=\frac{1}{8}$ supersymmetries. 
In this case the projection for the  Killing spinor is given by
\begin{align}
\label{cy4susy2}
\Gamma^{12910}\epsilon=\Gamma^{91078}\epsilon
=\Gamma^{7856}\epsilon=-\epsilon.
\end{align}
\item $CY_{3}$

In this case the geometry is given by $\mathbb{R}^{1,4}\times CY_{3}$. 
This decomposes the $Spin(6)$ into $SU(3)$ 
and correspondingly spinor representation decomposes as
\begin{align}
\label{cy3susy}
\bm{4}&=\bm{3}_{-}\oplus \bm{1}_{3}\nonumber\\
\overline{\bm{4}}&=\bm{3}_{+}\oplus \bm{1}_{-3}.
\end{align}
The appearance of two singlets from eight components  
means that there are $\frac{2}{8}=\frac{1}{4}$ supersymmetries in the 
product space $\mathbb{R}^{1,4}\times CY_{3}$. 
Therefore the Killing spinor can be defined by the projection
\begin{align}
\label{cy3susy2}
\Gamma^{12910}\epsilon=\Gamma^{91078}\epsilon=-\epsilon.
\end{align}

\item K3

For this case the geometry is the product space 
$\mathbb{R}^{1,6}\times$ K3. 
The decomposition of $Spin(4)$ into $SU(2)\times SU(2)$ 
gives rise to that of the spinor representation 
\begin{align}
\label{cy2susy}
\bm{2}&=(\bm{2},\bm{1})\nonumber\\
\bm{2}'&=(\bm{1},\bm{2}).
\end{align}
The presence of two singlets under one part of the $SU(2)$ implies that 
there are $\frac{2}{4}=\frac{1}{2}$ supersymmetries in the 
geometry $\mathbb{R}^{1,6}\times$ K3. 
The corresponding Killing spinors satisfy the projection 
\begin{align}
\label{cy2susy2}
\Gamma^{12910}\epsilon=-\epsilon.
\end{align}
\end{enumerate}

\subsubsection{Calibration and supersymmetric cycle}
\label{seccurv1b}
Now consider the situation where 
the M2-branes wrapping a Riemann surface $\Sigma_{g}$ 
propagate in a Calabi-Yau space without back reaction. 
To preserve supersymmetry on the world-volume, 
$\Sigma_{g}$ turns out to be a calibrated two-cycle, 
i.e. holomorphic curve of a Calabi-Yau manifold. 
To see this let us briefly review the background material 
concerning a calibration. 
In general a calibration on a special holonomy manifold $X$ 
is a differential $p$-form $\varphi$ obeying \cite{MR666108}
\begin{align}
\label{calibdef1}
d\varphi&=0\\
\label{calibdef2}
\varphi|_{\mathcal{C}_{p}}&\le \mathrm{Vol}|_{\mathcal{C}_{p}}, \ \ \
 \forall \mathcal{C}_{p}
\end{align}
where $\mathcal{C}_{p}$ is any $p$-cycle in $X$ 
and $\mathrm{Vol}$ is the volume form on the cycle induced from the
metric on $X$. 
A $p$-cycle $\Sigma$ is said to be calibrated by $\varphi$ 
if it satisfies 
\begin{align}
\label{calibdef3}
\varphi|_{\Sigma}=\mathrm{Vol}|_{\Sigma}.
\end{align}
We remark that a calibrated submanifold is a minimal surface 
in their homology class because
\begin{align}
\mathrm{Vol}(\Sigma)=
\int_{\Sigma}\varphi=\int_{M_{p+1}}d\varphi 
+\int_{\Sigma'}\varphi=\int_{\Sigma'}\varphi
\le \mathrm{Vol}(\Sigma')
\end{align}
where $\Sigma'$ is another $p$-cycle in the same homology class 
such that $\partial M_{p+1}=\Sigma-\Sigma'$.

It is known that Calabi-Yau $(n+1)$-folds admit two calibrations; 
the K\"{a}hler form $J$ and the holomorphic $(n+1,0)$-form $\Omega$.
One can construct calibrations as bilinear forms of spinors 
\cite{MR1208563,MR1045637}
\begin{align}
\label{calib4}
J_{MN}&=i\epsilon^{\dag}\Gamma_{MN}\epsilon\\
\Omega_{M_{1}\cdots M_{n+1}}
&=\epsilon^{T}\Gamma_{M_{1}\cdots M_{2(n+1)}}\epsilon
\end{align}

Now we consider the condition 
so that a bosonic configuration of membranes is supersymmetric. 
Since one can always add a second probe brane 
without breaking supersymmetry if it is wrapped on 
the supersymmetric cycle which the original probe brane is wrapping, 
a simple way to find such condition is to analyze an effective
world-volume action of a single membrane \cite{Becker:1995kb}. 
The action for a supermembrane coupled to $d=11$ supergravity is given
by \cite{Bergshoeff:1987cm}
\begin{align}
\label{memac01}
S=\int d^{3}x\Biggl[&
\frac12 \sqrt{-h}h^{\mu\nu}
\partial_{\mu}X^{M}\partial_{\nu}X^{N}g_{MN}
-\frac12 \sqrt{-h}\nonumber\\
&-i\sqrt{-h}h^{\mu\nu}
\overline{\Theta}\Gamma_{\mu}\nabla_{\nu}\Theta
+\frac16 \epsilon^{\mu\nu\lambda}C_{MNP}
\partial_{\mu}X^{M}\partial_{\nu}X^{N}\partial_{\lambda}X^{P}
\Biggr]
\end{align}
where $h_{\mu\nu},\mu,\nu=0,1,2$ is the metric of the world-volume, 
$h=\mathrm{det}(h_{\mu\nu})$, 
$g_{MN},M=0,1,\cdots,10$ is the $d=11$ space-time metric.  
$X^{M}$ is a space-time coordinate 
and $\Theta$ is a fermionic space-time coordinate. 
$C_{MNP}$ is a three-form gauge field, 
which is now taken to be zero in our background geometries.  
The action (\ref{memac01}) is invariant under 
the rigid supersymmetry transformations
\begin{align}
\label{memsusy01}
\delta_{\epsilon}X^{M}&=i\overline{\epsilon}\Gamma^{M}\Theta\\
\label{memsusy02}
\delta_{\epsilon}\Theta&=\epsilon
\end{align}
where $\epsilon$ is a constant anti-commuting eleven-dimensional spinor. 
Also the action (\ref{memsusy01}) has a local fermionic symmetry, 
called $\kappa$-symmetry. 
The $\kappa$-symmetry transformation is given by
\begin{align}
\label{kappasym1}
\delta_{\kappa} X^{M}&=
2i\overline{\Theta}\Gamma^{M}P_{+}\kappa(x)\\
\label{kappasym2}
\delta_{\kappa} \Theta&=
2P_{+}\kappa(x)
\end{align}
where $\kappa(x)$ is a $d=11$ spinor and 
the matrix 
\begin{align}
\label{pro11a}
P_{\pm}=
\frac12 
\left(
1\pm\frac{1}{6\sqrt{-h}}\epsilon^{\mu\nu\lambda}
\partial_{\mu}X^{M}
\partial_{\nu}X^{N}
\partial_{\lambda}X^{P}
\Gamma_{MNP}
\right)
\end{align}
is a projection operator satisfying
\begin{align}
\label{pro11b}
P_{\pm}^{2}=1,\ \ \ \ \ 
P_{+}P_{-}=0,\ \ \ \ \ 
P_{+}+P_{-}=1.
\end{align}

To extract the physical degrees of freedom, 
we must choose the suitable gauge 
that fixes the local world-volume reparametrization 
and the local $\kappa$-symmetry. 
Let us fix the reparametrization by choosing $x^{0}=X^{0}$. 
Then the projection operator (\ref{pro11a}) can be expressed as
\begin{align}
\label{pro11c}
P_{\pm}&=\frac12 
\left(
1\pm \Gamma
\right)
\end{align}
where 
\begin{align}
\Gamma:=\frac{1}{2\sqrt{\mathrm{det} (h_{\Sigma ij})}}\Gamma^{0}\epsilon^{ij}
\partial_{i}X^{M}\partial_{j}X^{N}\Gamma_{MN}.
\end{align}
Here $h_{\Sigma ij}, i,j=1,2$ is the metric of the Riemann surface wrapped 
by the M2-brane and $\sqrt{\mathrm{det}(h_{\Sigma ij})}$ is the area 
of the surface. 
As a next step we want to fix the local $\kappa$-symmetry on the
world-volume. 
In order for a bosonic world-volume configuration to be supersymmetric, 
the global supersymmetry transformations (\ref{memsusy02}) 
need to be compensated for by the $\kappa$-symmetry transformations 
(\ref{kappasym2}) 
\begin{align}
\label{memsusy03}
\left(
\delta_{\epsilon}+\delta_{\kappa}
\right)\Theta=\epsilon+2P_{+}\kappa(x)=0.
\end{align}
Acting $P_{-}$ on both sides we find that 
\begin{align}
\label{memsusy04}
P_{-}\epsilon=
\frac{1-\Gamma}{2}\epsilon
=0.
\end{align}
Therefore the supersymmetry preserved by the M2-branes 
is given by the Killing spinor $\epsilon$ which obeys the projection 
(\ref{memsusy03}). 
Noting that $\Gamma^{2}=1$ and $\Gamma^{\dag}=\Gamma$, 
we find that
\begin{align}
\label{memsusy04}
\epsilon^{\dag}
\frac{1-\Gamma}{2}
\epsilon
=\epsilon^{\dag}\frac{(1-\Gamma)(1-\Gamma)}{2}\epsilon
=\left|\frac{1-\Gamma}{\sqrt{2}}\epsilon\right|^{2}\ge 0. 
\end{align}
By normalizing the Killing spinors such that $\epsilon^{\dag}\epsilon=1$, 
the inequality (\ref{memsusy04}) can be rewritten as
\begin{align}
\label{memsusy05}
\mathrm{Vol}(\Sigma_{g})\ge \varphi
\end{align}
where $\mathrm{Vol}(\Sigma_{g})=\sqrt{\mathrm{det}(h_{\Sigma ij})}$ is 
the area of the Riemann surface and $\varphi$ is the differential
two-form defined by
\begin{align}
\label{memsusy06}
\varphi=-\frac12
\left( 
\overline{\epsilon}\Gamma_{MN}\epsilon 
\right)
dX^{M}\wedge dX^{N}.
\end{align}
Hence the two-form (\ref{memsusy06}) 
satisfies the condition (\ref{calibdef2}) for the calibration 
and has the bilinear expression for K\"{a}hler calibration $J$ 
(see (\ref{calib4})). 
Moreover it can be shown that 
the two-form (\ref{memsusy06}) obeys the other required condition
(\ref{calibdef1}) for the calibration 
by using the supersymmetry algebra \cite{Gutowski:1999tu}. 
Therefore we can conclude that the two-form (\ref{memsusy06}) is 
a K\"{a}hler calibration and that 
the supersymmetric two-cycle $\Sigma_{g}$ 
wrapped by the M2-branes is a calibrated two-cycle, 
i.e. a holomorphic curve. 
Notice that 
(\ref{memsusy03}) is precisely 
the chirality condition $\Gamma^{012}\epsilon=0$ imposed 
on the supersymmetry parameters in the BLG-model (see
(\ref{blgchiralmtx})).

At this stage we are ready to count the number of preserved
supersymmetries in our M2-brane configurations 
by combining the two different types of projections; 
the projections (\ref{cy5susy2}), (\ref{cy4susy2}), (\ref{cy3susy2}) and
(\ref{cy2susy2}) 
for the background Calabi-Yau manifolds 
and the projection (\ref{memsusy03}) (or (\ref{blgchiralmtx})) 
for the membranes wrapped 
around a calibrated two-cycle $\Sigma_{g}$. 
In most of the cases wrapped branes break half of the supersymmetries 
preserved by the special holonomy manifolds according to the additional
projection for the branes wrapping calibrated cycles. 
However, for the Calabi-Yau 5-fold 
the projection condition (\ref{memsusy03}) 
for the M2-branes does not give rise to 
a further constraint on the surviving two Killing spinors. 
This implies that  M2-branes can wrap a holomorphic curve 
in a Calabi-Yau 5-fold without breaking down the supersymmetry. 
The amounts of preserved supersymmetries by the M2-branes wrapping 
holomorphic curves $\Sigma_{g}$ in Calabi-Yau spaces  
are summarized as
\begin{align}
\label{calibeq}
\mathcal{N}=
\begin{cases}
\ 8&\textrm{for $\Sigma_{g}\subset$ K3}\cr
\ 4&\textrm{for $\Sigma_{g}\subset CY_{3}$}\cr
\ 2&\textrm{for $\Sigma_{g}\subset CY_{4}$}\cr
\ 2&\textrm{for $\Sigma_{g}\subset CY_{5}$}.\cr
\end{cases}
\end{align}
%
%
%
%
%
%
%
%
%
Upon the dimensional reduction to $\mathbb{R}$, 
the arising quantum mechanics on $\mathbb{R}$ will have 
the same number of supersymmetries.

\subsection{Topological twisting}
\label{seccurv2}
In general a quantum field theory on the curved M2-branes 
interacts with gravity, however, 
it is also possible to get a supersymmetric quantum field theory 
on $\mathbb{R}\times\Sigma_{g}$ by taking the appropriate decoupling limit 
$l_{p}\rightarrow 0$ while keeping the volume of $\Sigma_{g}$ and that
of $X$ fixed. 
In order to derive such low-energy effective theories on the curved 
world-volume, 
we recall how the BLG-model describes the dynamics 
of the flat M2-branes. 
In the BLG-model the fields and supercharges transform under 
$SO(2)_{E}\times SO(8)_{R}$ as
\begin{align}
\label{blgfieldrep1}
X_{a}^{I}& : {\bm{8}_{v}}_{0}\nonumber \\
\Psi_{a}& : \bm{8}_{c+}\oplus \bm{8}_{c-}\nonumber \\
\epsilon& : \bm{8}_{s+}\oplus \bm{8}_{s-}.
\end{align}
The eight scalar fields $X^{I}$'s transform as the vector
representations of the R-symmetry $SO(8)_{R}$ which represents
 the rotational group of the transverse space of the M2-branes. 
In other words, they are sections of the normal bundle, 
which is trivial in this case. 
However, corresponding to the geometry given in (\ref{m2geo}), 
now the tangent bundle $T_{X}$ of the ambient Calabi-Yau manifold $X$ 
is decomposed as
\begin{equation}
 T_{X}=T_{\Sigma}\oplus N_{\Sigma}
\end{equation}
where $T_{\Sigma}$ is the tangent bundle over the Riemann surface
$\Sigma_{g}$ and $N_{\Sigma}$ is the
normal bundle over the surface. 
Therefore we need to take into account the existence of 
the non-trivial normal bundle of calibrated cycles and 
to introduce new dynamical variables instead of the original scalar fields. 
These transitions from scalars, 
i.e. trivial normal bundle to the
non-trivial normal bundles are intimately connected with
the way in which the field theory on $\mathbb{R}\times \Sigma_{g}$ 
realizes supersymmetry. 
Along with the coupling to the curvature on the Riemann surface, 
there now exists a coupling to an external $SO(2n)$ gauge group, 
the R-symmetry background. 
Thus one can preserve supersymmetry on the holomorphic Riemann surface 
by choosing the $SO(2)$ Abelian background 
from the $SO(2n)$ appropriately.

There is a beautiful observation that 
such an effective description for curved branes can be obtained   
by topological twisting \cite{Bershadsky:1995qy}. 
Here we attempt to twist the BLG-model to obtain 
the low-energy descriptions for the curved M2-branes 
\footnote{
For the ABJM-model 
the geometric meaning of the topological twisting is less clear 
because the classical $SU(4)_{R}$ R-symmetry reflects the orbifolds. 
In this paper we will focus on the BLG-model.
}.
Schematically topological twisting procedure can be achieved by replacing the
original Euclidean rotational group $SO(2)_{E}$ on the Riemann surface by a
different subgroup $SO(2)'_{E}$ of $SO(2)_{E}\times SO(8)_{R}$. 
Although there are many possible ways to pick such subgroups,  
here we will consider the following decomposition
\begin{align}
\label{splitso8}
 SO(8)\supset
& SO(8-2n)\times SO(2n) \nonumber \\
\supset
& SO(8-2n)\times SO(2)_{1}\times \cdots \times SO(2)_{n}.
\end{align}
The $SO(8-2n)$ is a rotational group of the Euclidean space
perpendicular to the Riemann surface, 
while the $SO(2)_{i}$ are
diagonal subgroups of the external $SO(2n)$ gauge group. 
The meaning of this decomposition is that 
the Calabi-Yau manifold $X$ enjoys the decomposable line bundles as the
form 
\begin{align}
\label{cyconst1}
X=\mathcal{L}_{1}\oplus
\cdots \oplus
\mathcal{L}_{n}
\rightarrow \Sigma_{g}.
\end{align} 
Under the decomposition (\ref{splitso8}), 
the R-charges for $\bm{8}_{v}$, $\bm{8}_{s}$
and $\bm{8}_{c}$ are determined as follows:

\begin{enumerate}
 \item $SO(8)\supset SO(6)\times SO(2)_{1}$
\begin{align}
\label{k3twist}
\bm{8}_{v}=&\bm{6}_{0}\oplus\bm{1}_{2}\oplus\bm{1}_{-2} \nonumber \\
\bm{8}_{s}=&\bm{4}_{+}\oplus\overline{\bm{4}}_{-} \nonumber \\
\bm{8}_{c}=&\bm{4}_{-}\oplus\overline{\bm{4}}_{+}.
\end{align}

\item $SO(8)
\supset SO(4)\times SO(2)_{1}\times SO(2)_{2}$
\begin{align}
\label{cy3twist1}
\bm{8}_{v}
=&\bm{4}_{00}\oplus \bm{1}_{02}\oplus \bm{1}_{0-2}\oplus
 \bm{1}_{20}\oplus \bm{1}_{-20}\nonumber \\
\bm{8}_{s}
=&\bm{2}_{++}\oplus \bm{2}'_{+-}\oplus \bm{2}_{--}\oplus
 \bm{2}_{-+}'\nonumber \\
\bm{8}_{c}
=&\bm{2}_{-+}\oplus \bm{2}'_{--}\oplus \bm{2}_{+-}\oplus \bm{2}_{++}'.
\end{align}

\item $SO(8)
\supset SO(2)\times SO(2)_{1}\times SO(2)_{2}\times
      SO(2)_{3}$
\begin{align}
\label{cy4twist1}
\bm{8}_{v}
=&
\bm{2}_{000}\oplus\bm{1}_{002}\oplus\bm{1}_{00-2}
\oplus\bm{1}_{020}\oplus\bm{1}_{0-20}
\oplus\bm{1}_{200}\oplus\bm{1}_{-200}
\nonumber \\
\bm{8}_{s}
=&
\bm{1}_{+++}\oplus\bm{1}_{++-}
\oplus\bm{1}_{+--}\oplus\bm{1}_{+-+}
\oplus\bm{1}_{--+}\oplus\bm{1}_{---}
\oplus\bm{1}_{-+-}\oplus\bm{1}_{-++}
\nonumber \\
\bm{8}_{c}
=&
\bm{1}_{-++}\oplus\bm{1}_{-+-}
\oplus\bm{1}_{---}\oplus\bm{1}_{--+}
\oplus\bm{1}_{+-+}\oplus\bm{1}_{+--}
\oplus\bm{1}_{++-}\oplus\bm{1}_{+++}.
\end{align}

\item $SO(8)
\supset SO(2)_{1}\times SO(2)_{2}\times
      SO(2)_{3}\times SO(2)_{4}$
\begin{align}
\label{cy5twist1}
\bm{8}_{v}
=&
\bm{1}_{0002}\oplus\bm{1}_{000-2}\oplus\bm{1}_{0020}\oplus\bm{1}_{00-20}
\oplus\bm{1}_{0200}\oplus\bm{1}_{0-200}
\oplus\bm{1}_{2000}\oplus\bm{1}_{-2000}
\nonumber \\
\bm{8}_{s}
=&
\bm{1}_{++++}\oplus\bm{1}_{++--}
\oplus\bm{1}_{+--+}\oplus\bm{1}_{+-+-}
\oplus\bm{1}_{--++}\oplus\bm{1}_{----}
\oplus\bm{1}_{-+-+}\oplus\bm{1}_{-++-}
\nonumber \\
\bm{8}_{c}
=&
\bm{1}_{-+++}\oplus\bm{1}_{-+--}
\oplus\bm{1}_{---+}\oplus\bm{1}_{--+-}
\oplus\bm{1}_{+-++}\oplus\bm{1}_{+---}
\oplus\bm{1}_{++-+}\oplus\bm{1}_{+++-}.
\end{align}

\end{enumerate}
With one of the decompositions (\ref{k3twist})-(\ref{cy5twist1}), 
we can now define a new generator $s'$, 
i.e. the $SO(2)_{E}'$ charge by
\begin{align}
\label{twistcharge1}
s':=s-\sum_{i=1}^{n}a_{i}T_{i}.
\end{align}
Here $s$ denotes a generator of the original rotational group
$SO(2)_{E}$, $T_{i}$ represents a 
generator of the subgroup $SO(2)_{i}$ diagonally embedded
in the external gauge group $SO(2n)$ 
and $a_{i}$'s are 
the constant parameters characterizing the twisting procedures. 
From now on we normalize these charges $s'$, $s$ and $T_{i}$ 
such that they are twice as the usual spin on the Riemann
surface. 
Since $a_{i}$'s are related to 
the degrees of the line bundles $\mathcal{L}_{i}$'s as 
\begin{align}
\label{bdldeg}
\deg (\mathcal{L}_{i})=
\begin{cases}
2|g-1|a_{i}& \textrm{for} \ \ g\neq 0\cr
a_{i}& \textrm{for} \ \ g=0\cr
\end{cases}
\end{align}
and the degrees coincide with the first Chern class, 
the conditions that $X$ is Calabi-Yau are given by 
\begin{align}
\label{cycond1}
\sum_{i=1}^{n}a_{i}=
\begin{cases}
-1& \textrm{for} \ \ g=0\cr
0& \textrm{for} \ \ g=1\cr
1& \textrm{for} \ \ g>1\cr 
\end{cases}.
\end{align}
Note that the Calabi-Yau conditions (\ref{cycond1}) 
simultaneously ensure the existence of the covariant constant spinors 
in the twisted theories. 
One can easily check that 
the topological twists underlying the decompositions 
(\ref{k3twist}), (\ref{cy3twist1}), (\ref{cy4twist1}) and 
(\ref{cy5twist1}) preserve $8$, $4$, $2$ and $2$ 
supersymmetries as we expect for K3, $CY_{3}$, $CY_{4}$ and $CY_{5}$.

Therefore given the decomposable line bundle structures 
of the Calabi-Yau manifolds (\ref{cyconst1}), 
we can determine 
the topological twisting procedure 
from the two conditions (\ref{bdldeg}) and (\ref{cycond1}). 
For a K3 surface, i.e. for 
$a_{2}=a_{3}=a_{4}=0$, 
the local geometry is $T^{*}\Sigma_{g}$ and 
a single twisting parameter $a_{1}$ is uniquely determined by 
the Calabi-Yau condition up to the orientation.  
For other Calabi-Yau spaces 
the Calabi-Yau conditions are not so powerful and 
there are infinitely many ways of the twisting 
characterized by $a_{i}$, or the degrees of the line bundles.

\section{SCQM from M2-branes in a K3 surface}
\label{cy2sec}
Let us study the membranes wrapping a curved Riemann surface of genus $g>1$
embedded in a K3 surface. 
In order to preserve supersymmetry one should carry out the 
topological twisting utilizing the decomposition (\ref{k3twist}). 
Requiring the existence of covariant constant spinors, 
the twisting procedure can be uniquely determined 
since the external gauge field is nothing but an $SO(2)$ Abelian
background in this case. 
Note that the twisting for $\Sigma_{g}=\mathbb{P}^{1}$ can be 
realized just by the orientation reversal.

Under $SO(2)_{E}\times SO(8)_{R}\rightarrow SO(2)_{E}'\times SO(6)_{R}$, 
the twisted field theory with $g>1$ 
is characterized by the following representations
\begin{align}
\label{k3content}
&X^{I}:\bm{8}_{v0}\rightarrow\bm{6}_{0}\oplus \bm{1}_{2}\oplus
 \bm{1}_{-2} \nonumber \\
&\epsilon:\bm{8}_{s+}\oplus \bm{8}_{s-}\rightarrow\bm{4}_{0}\oplus\overline{\bm{4}}_{2}
\oplus\bm{4}_{-2}\oplus \overline{\bm{4}}_{0}\nonumber \\
&\Psi:\bm{8}_{c+}\oplus\bm{8}_{c-}\rightarrow
\bm{4}_{2}\oplus \overline{\bm{4}}_{0}
\oplus \bm{4}_{0}\oplus \overline{\bm{4}}_{-2}.
\end{align}
Therefore the bosonic field content is 
six scalar fields 
$\phi^{I}$ transforming as 
$\bm{6}_{0}$
and one-forms 
$\Phi_{z}$, $\Phi_{\overline{z}}$ transforming as 
$\bm{1}_{2}\oplus \bm{1}_{-2}$. 
The fermionic field content is 
eight scalar fields 
$\psi,\tilde{\lambda}$ 
as $\bm{4}_{0}\oplus\overline{\bm{4}}_{0}$
and 
one-forms $\Psi_{z}, \tilde{\Psi}_{\overline{z}}$ 
as $\bm{4}_{2}\oplus \overline{\bm{4}}_{-2}$. 
The supersymmetry parameters are 
eight scalars 
$\epsilon,\tilde{\epsilon}$ 
as $\bm{4}_{0}\oplus \overline{\bm{4}}_{0}$ 
and 
one-forms $\tilde{\epsilon}_{z}, \epsilon_{\overline{z}}$ 
as $\overline{\bm{4}}_{2}\oplus \bm{4}_{-2}$. 
Here and hereafter we distinguish $\bm{4}$ and $\overline{\bm{4}}$ 
in terms of tildes over the fermionic objects.

We should note that 
there are six bosonic scalar fields 
and eight fermionic scalar charges in the twisted theory. 
Since a Riemann surface is 
a real two-dimensional manifold and there are six scalar fields, 
the theory should describe the circumstance where the 
two-cycle lives in a $2+(8-6)=4$-dimensional curved manifold $X$. 
The existence of eight scalar supercharges indicates that 
the four-manifold preserves $\frac{8}{16}=\frac12$ of the supersymmetries. 
This is the case where a holomorphic Riemann surface $\Sigma_{g}$ is
embedded in a K3 surface.

Locally the K3 geometry is the cotangent bundle $T^{*}\Sigma_{g}$.
The remaining two scalar fields combine to yield  
one-forms on the Riemann surface. 
They represent the motion of the M2-branes 
along the non-trivial normal bundle $N_{\Sigma}$ 
over the Riemann surface inside the K3 surface. 
Under the $SO(6)$ rotational group 
of the six uncompactified dimensions, 
the six scalars transform as vector representations $\bm{6}_{v}$ 
and the one-forms are just singlets. 
We take the eleven-dimensional space-time configuration as
\begin{equation}
\label{k3}
 \begin{array}{cccccccccccc}
&0&1&2&3&4&5&6&7&8&9&10\\
\textrm{K3}&\times&\circ&\circ&\times&\times&\times&\times&\times&\times&\circ&\circ\\
\textrm{M2}&\circ&\circ&\circ&\times&\times&\times&\times&\times&\times&\times&\times \\
\Sigma_{g}&\times&\circ&\circ&\times&\times&\times&\times&\times&\times&\times&\times\\
\end{array}
\end{equation}
where $\circ$ denotes the direction 
in which the geometrical objects extend, while  
$\times$ denotes the direction in which they localize. 
Note that the projection (\ref{cy2susy2}) for the K3 surface encodes 
the configuration (\ref{k3}). 
The world-volume of the M2-branes extend to 
a time direction $x^{0}$ and spacial directions $x^{1}, x^{2}$.  
The spacial directions $x^{1}$, $x^{2}$ 
are tangent to the compact Riemann surface in the K3 surface. 
The normal geometry of the M2-branes is divided into two parts;    
one is the normal bundle $N_{\Sigma}$ inside the K3 surface, 
extending to two directions $x^{9}$, $x^{10}$ and 
the other is the flat Euclidean space transverse to the K3 surface, labeled
by $x^{3},\cdots, x^{8}$.

\subsection{Twisted theory}
\label{cy2sec1}
Firstly our space-time configuration 
(\ref{k3}) breaks down the space-time symmetry group 
$SO(1,10)$ to $SO(2)_{E}\times SO(6)_{R}\times SO(2)_{1}$. 
So the $SO(1,10)$ gamma matrix can be decomposed as
\begin{equation}
\label{mtx1}
\begin{cases}
 \Gamma^{\mu}=\gamma^{\mu}\otimes
  \hat{\Gamma}^{7}
\otimes \sigma_{2}
&\mu=0,1,2 \cr
 \Gamma^{I+2}=\mathbb{I}_{2}\otimes
 \hat{\Gamma}^{I} \otimes 
\sigma_{2}&I=1,\cdots, 6\cr
 \Gamma^{i+8}=\mathbb{I}_{2}\otimes
 \mathbb{I}_{8}\otimes
 \gamma^{i}&i=1,2 \cr
\end{cases}
\end{equation}
where
 $\hat{\Gamma}^{I}$ is the $SO(6)$ gamma matrix obeying
\begin{align}
\{\hat{\Gamma}^{I},\hat{\Gamma}^{J}\}=2\delta^{IJ},\ \ \ 
(\hat{\Gamma}^{I})^{\dag}=\Gamma^{I}
\end{align}
\begin{equation}
\label{gamma7}
\hat{\Gamma}^{7}
=-i\hat{\Gamma}^{12\cdots 6}=\left(
\begin{array}{cc}
\mathbb{I}_{4}&0\\
0&-\mathbb{I}_{4}\\
\end{array}
\right).
\end{equation} 
Similarly the $SO(1,10)$ charge conjugation matrix $\mathcal{C}$ is
expressed as
\begin{equation}
\label{ccmtx}
 \mathcal{C}=\epsilon \otimes \hat{C} \otimes \epsilon
\end{equation}
where $\epsilon:=i\sigma_{2}$ is introduced as the charge conjugation
matrix with the relations
\begin{equation}
\epsilon^{T}=-\epsilon,\ \ \
\epsilon\gamma^{\mu}\epsilon^{-1}=-(\gamma^{\mu})^{T} 
\end{equation}
while $\hat{C}$ is the $SO(6)$ charge conjugation matrix satisfying
\begin{align}
\label{ccmtx1}
\hat{C}^{T}=-\hat{C},\ \ \ \ \ \ \hat{C}\hat{\Gamma}^{I}\hat{C}^{-1}=(\hat{\Gamma}^{I})^{T},\ \ \ \ \ \ 
\hat{C}\hat{\Gamma}^{7}\hat{C}^{-1}=-(\hat{\Gamma}^{7})^{T}.
\end{align}
Under the decomposition (\ref{mtx1}), 
the $SO(8)$ chiral matrix becomes
\begin{equation}
\label{chiralmtx}
\Gamma^{012}=
\Gamma^{34\cdots 10}=
\mathbb{I}_{2}\otimes \hat{\Gamma}^{7}\otimes \sigma_{2}.
\end{equation}

For the twisted bosonic fields we set
\begin{align}
\label{new1}
\phi^{I}&:=X^{I+2}\\
\label{new2}
\Phi_{z}&:=\frac{1}{\sqrt{2}}(X^{9}-iX^{10}), 
&\Phi_{\overline{z}}&:=\frac{1}{\sqrt{2}}(X^{9}+iX^{10})\\
\label{new3}
A_{z}&:=\frac{1}{\sqrt{2}}(A_{1}-iA_{2}), 
&A_{\overline{z}}&:=\frac{1}{\sqrt{2}}(A_{1}+iA_{2})
\end{align}
where the bosonic scalar fields $\phi^{I}$'s transform as 
 the vector representations $\bm{6}_{v}$ of the $SO(6)$ global
symmetry and the indices $I=1,\cdots,6$ label 
the flat transverse directions. 
The bosonic one-froms, $\Phi_{z}$ and $\Phi_{\overline{z}}$ are 
the $SO(6)$-singlets 
and they describe the motion in the 
normal geometry $N_{\Sigma}$ of the Riemann
surface inside the K3 surface. 
These Higgs fields $\phi^{I}, \Phi_{z}$ and $\Phi_{\overline{z}}$ 
are the 3-algebra valued.

Now consider the twisted fermionic objects. 
Primitively 
the fermionic fields $\Psi$ are 
$SL(2,\mathbb{R})$ spinors that transform as the spinor 
representations $\bm{8}_{c}$ of the $SO(8)_{R}$ R-symmetry. 
As seen from (\ref{k3content}), 
under the decomposition  
$Spin(1,10)$ $\rightarrow$ $Spin(2)\times Spin(6)\times Spin(2)$,  
fermionic fields $\Psi$ are split into the representations 
$\bm{4}_{2}$, $\overline{\bm{4}}_{0}$, 
$\bm{4}_{0}$ and $\overline{\bm{4}}_{-2}$, 
whose component fields are denoted by 
$\Psi_{z}$, $\tilde{\lambda}$, $\psi$ and
$\tilde{\Psi}_{\overline{z}}$ respectively. 
Accordingly they can be expanded as
\begin{align}
\label{new4}
\Psi_{A}^{\alpha\beta}
=\frac{i}{\sqrt{2}}
\psi_{A}(\gamma_{+}\epsilon^{-1})^{\alpha\beta}
+i\tilde{\Psi}_{\overline{z}A}
(\gamma^{\overline{z}}\epsilon^{-1})^{\alpha\beta}
-\frac{i}{\sqrt{2}}\tilde{\lambda}_{A}
(\gamma_{-}\epsilon^{-1})^{\alpha\beta}
-i\Psi_{zA}(\gamma^{z}\epsilon^{-1})^{\alpha\beta}
\end{align}
where the three indices $\alpha,A$ and $\beta$ denote 
the $SO(2)_{E}$ spinor,
the $SO(6)_{R}$ spinor 
and the $SO(2)_{1}$ spinor respectively. 
Here we have introduced the matrices $\gamma_{\pm}, \gamma^{z}$ and
$\gamma^{\overline{z}}$ defined by 
\begin{align}
\label{gammapm}
&\gamma_{+}:=\frac{1}{\sqrt{2}}(\mathbb{I}_{2}+\sigma_{2}),\ \ \ \ \ 
\gamma_{-}:=\frac{1}{\sqrt{2}}(\mathbb{I}_{2}-\sigma_{2})\\
\label{gammaz1}
&\gamma^{z}
:=\frac{1}{\sqrt{2}}
(\gamma^{1}+i\gamma^{2})=\frac{1}{\sqrt{2}}\left(
\begin{array}{cc}
i&1\\
1&-i\\
\end{array}
\right) \\
\label{gammaz2}
&\gamma^{\overline{z}}
:=\frac{1}{\sqrt{2}}(\gamma^{1}-i\gamma^{2})
=\frac{1}{\sqrt{2}}\left(
\begin{array}{cc}
-i&1\\
1&i\\
\end{array}
\right).
\end{align}
As seen from (\ref{new4}), 
the above matrices enable us to carry out the topological twisting, 
or in other words the identification of the index $\alpha$ 
with the index $\beta$. 
The matrices $\gamma_{+}$ and $\gamma^{\overline{z}}$ are associated with 
the conjugate spinor representations $\bm{8}_{c-}$ 
and yield $\bm{4}_{0}$ and $\overline{\bm{4}}_{-2}$,  
while the other pair of matrices $\gamma_{-}$ and $\gamma^{z}$ are 
associated with $\bm{8}_{c+}$ 
and give rise to $\bm{4}_{2}$ and $\overline{\bm{4}}_{0}$. 
Together with the decomposition (\ref{chiralmtx}) 
and the chirality condition (\ref{blgchiral}) for $\Psi$, 
one can check that the expansion (\ref{new4}) 
leads to the relations; 
$\hat{\Gamma}^{7}\psi=\psi$, 
$\hat{\Gamma}^{7}\tilde{\Psi}_{\overline{z}}=-\tilde{\Psi}_{\overline{z}}$, 
$\hat{\Gamma}^{7}\tilde{\lambda}=-\tilde{\lambda}$ and
$\hat{\Gamma}^{7}\Psi_{z}=\Psi_{z}$. 
For the $\mathcal{A}_{4}$ algebra  
all of these fermionic fields are the fundamental representations 
of the $SO(4)$ gauge group. 
We define the conjugate of the $SO(6)$ spinors as
\begin{align}
&\overline{\psi}:=\psi^{T}\hat{C},\ \ \
 \overline{\tilde{\lambda}}:=\tilde{\lambda}^{T}\hat{C},\ \ \ 
\overline{\Psi}_{z}:=\Psi_{z}^{T}\hat{C},\ \ \ 
\overline{\tilde{\Psi}}_{\overline{z}}:=\tilde{\Psi}_{\overline{z}}^{T}\hat{C}.
\end{align}

Likewise, the supersymmetry parameters originally transform as the 
$SL(2,\mathbb{R})$ spinor representations of the rotational group 
of the world-volume and $\bm{8}_{s}$ of the $SO(8)$ R-symmetry in
the BLG-model, 
while in the twisted theory they reduce to the four distinct representations 
$\bm{4}_{0}$, $\overline{\bm{4}}_{2}$, 
$\bm{4}_{-2}$ and $\overline{\bm{4}}_{0}$.  
Thus we can write supersymmetry parameters as
\begin{align}
\label{new5}
\epsilon_{A}^{\alpha\beta}
=\frac{i}{\sqrt{2}}\tilde{\epsilon}_{A}
(\gamma_{+}\epsilon^{-1})^{\alpha\beta}
+i\epsilon_{\overline{z}A}
(\gamma^{\overline{z}}\epsilon^{-1})^{\alpha\beta}
-\frac{i}{\sqrt{2}}\epsilon_{A}
(\gamma_{-}\epsilon^{-1})^{\alpha\beta}
-i\tilde{\epsilon}_{zA}
(\gamma^{z}\epsilon^{-1})^{\alpha\beta}.
\end{align}
Here again the indices $\alpha$, $A$ and $\beta$ 
label $SO(2)_{E}$, $SO(6)_{R}$
and $SO(2)_{1}$ respectively. 
Since $\epsilon$ and $\tilde{\epsilon}$ 
are fermionic scalars on an arbitrary Riemann surface, 
they are identified with supercharges and hence 
the effective theory will be 
endowed with the corresponding eight supercharges.

In terms of the expressions (\ref{mtx1}), (\ref{new1}), (\ref{new2}),
(\ref{new3}) and (\ref{new4}), 
we obtain the twisted BLG Lagrangian
\begin{align}
\label{twistedlagrangian}
 \mathcal{L}
&=\frac12 (D_{0}\phi^{I},D_{0}\phi^{I})
-(D_{z}\phi^{I},D_{\overline{z}}\phi^{I})
+(D_{0}\Phi_{z},D_{0}\Phi_{\overline{z}})
-2(D_{z}\Phi_{\overline{w}},D_{\overline{z}}\Phi_{w})
\nonumber\\
&
+(\overline{\tilde{\lambda}},D_{0}\psi)
+(\overline{\Psi}_{z},D_{0}\tilde{\Psi}_{\overline{z}})
-(\overline{\tilde{\Psi}}_{\overline{z}},D_{0}\Psi_{z})
-2i(\overline{\tilde{\Psi}}_{\overline{z}},D_{z}\psi)
+2i(\overline{\tilde{\lambda}},D_{\overline{z}}\Psi_{z})
\nonumber \\
&
+\frac{i}{2}(\overline{\tilde{\lambda}}\hat{\Gamma}^{IJ},
[\phi^{I},\phi^{J},\psi])
-i(\overline{\tilde{\Psi}}_{\overline{z}}\hat{\Gamma}^{IJ},
[\phi^{I},\phi^{J},\Psi_{z}])
\nonumber \\
&
+2i(\overline{\psi}\hat{\Gamma}^{I},
[\Phi_{\overline{z}},\phi^{I},\Psi_{z}])
-2i(\overline{\tilde{\lambda}}\hat{\Gamma}^{I},
[\Phi_{z},\phi^{I},\tilde{\Psi}_{\overline{z}}])
\nonumber \\
&
+i(\overline{\tilde{\lambda}},
[\Phi_{z},\Phi_{\overline{z}},\psi])
-2i(\overline{\tilde{\Psi}}_{\overline{w}},
[\Phi_{z},\Phi_{\overline{z}},\Psi_{w}])
\nonumber \\
&
-\frac{1}{12}
\left(
[\phi^{I},\phi^{J},\phi^{K}],
[\phi^{I},\phi^{J},\phi^{K}]
\right)
-\frac12 \left(
[\Phi_{z},\phi^{I},\phi^{J}],
[\Phi_{\overline{z}},\phi^{I},\phi^{J}]
\right)
\nonumber \\
&
-\frac12 \left(
[\Phi_{z},\Phi_{w},\phi^{I}],
[\Phi_{\overline{z}},\Phi_{\overline{w}},\phi^{I}]
\right)
-\frac12 \left(
[\Phi_{z},\Phi_{\overline{w}},\phi^{I}],
[\Phi_{\overline{z}},\Phi_{w},\phi^{I}]
\right)
\nonumber \\
&
+\frac16 \left(
[\Phi_{z},\Phi_{w},\Phi_{v}],
[\Phi_{\overline{z}},\Phi_{\overline{w}},\Phi_{\overline{v}}]
\right)
+\frac12 \left(
[\Phi_{z},\Phi_{w},\Phi_{\overline{v}}],
[\Phi_{\overline{z}},\Phi_{\overline{w}},\Phi_{v}]
\right)
+\mathcal{L}_{\textrm{TCS}}.
\end{align}
Here we have introduced 
$(\ ,\ )$ 
as the trace form on the 
3-algebra introduced in (\ref{3algmet}) 
and we have defined the covariant derivatives 
$ D_{z}:=\frac{1}{\sqrt{2}}(D_{1}-iD_{2})$ and 
$D_{\overline{z}}:=\frac{1}{\sqrt{2}}(D_{1}+iD_{2})$.

The corresponding BRST transformations are given by
\begin{align}
\label{brstbos1}
\delta\phi^{I}_{a}
&=i\overline{\tilde{\epsilon}}\hat{\Gamma}^{I}\tilde{\lambda}_{a}
-i\overline{\epsilon}\hat{\Gamma}^{I}\psi_{a}\\
\label{brstbos2}
\delta\Phi_{za}
&=-i\overline{\tilde{\epsilon}}\Psi_{za}
\\
\label{brstbos3}
\delta\Phi_{\overline{z}a}
&=-i\overline{\epsilon}\tilde{\Psi}_{\overline{z}a}
\\
\label{brstfermi1}
\delta\psi_{a}
&=
iD_{0}\phi^{I}_{a}\hat{\Gamma}\tilde{\epsilon}
-2D_{\overline{z}}\Phi_{za}\epsilon
+\frac16
 [\phi^{I},\phi^{J},\phi^{K}]_{a}\hat{\Gamma}^{IJK}\tilde{\epsilon}
+[\Phi_{z},\Phi_{\overline{z}},\phi^{I}]_{a}
\hat{\Gamma}^{I}\tilde{\epsilon}
\\
\label{brstfermi2}
\delta\tilde{\lambda}_{a}
&=
iD_{0}\phi^{I}_{a}\hat{\Gamma}^{I}\epsilon
-2D_{z}\Phi_{\overline{z}a}\tilde{\epsilon}
-\frac16 
[\phi^{I},\phi^{J},\phi^{K}]_{a}\hat{\Gamma}^{IJK}\epsilon
+[\Phi_{z},\Phi_{\overline{z}},\phi^{I}]_{a}
\hat{\Gamma}^{I}\epsilon
\\
\label{brstfermi3}
\delta\Psi_{za}
&=
-D_{z}\phi^{I}\hat{\Gamma}^{I}\tilde{\epsilon}
-iD_{0}\Phi_{z}\epsilon
+\frac12 [\Phi_{z},\phi^{I},\phi^{J}]_{a}
\hat{\Gamma}^{IJ}\epsilon
+\frac13 [\Phi_{w},\Phi_{\overline{w}},\Phi_{z}]_{a}
\epsilon\\
\label{brstfermi4}
\delta\tilde{\Psi}_{\overline{z}a}
&=
D_{\overline{z}}\phi^{I}_{a}\hat{\Gamma}^{I}\epsilon
+iD_{0}\Phi_{\overline{z}a}\tilde{\epsilon}
+\frac12
 [\Phi_{\overline{z}},\phi^{I},\phi^{J}]_{a}
\hat{\Gamma}^{IJ}\tilde{\epsilon}
+\frac13 
[\Phi_{\overline{w}},\Phi_{w},\Phi_{\overline{z}}]_{a}
\tilde{\epsilon}\\
\label{brstbos4}
\delta\tilde{A}_{0 a}^{b}
&=-\overline{\epsilon}\hat{\Gamma}^{I}\phi_{c}^{I}\psi_{d}{f^{cdb}}_{a}
-\overline{\tilde{\epsilon}}\hat{\Gamma}^{I}\phi_{c}^{I}
\tilde{\lambda}_{d}{f^{cdb}}_{a}
-2\overline{\epsilon}\Phi_{zc}\tilde{\Psi}_{\overline{z}d}{f^{cdb}}_{a}
+2\overline{\tilde{\epsilon}}\Phi_{\overline{z}c}\Psi_{zd}{f^{cdb}}_{a}\\
\label{brstbos5}
\delta\tilde{A}^{b}_{za}
&=2i\overline{\epsilon}\hat{\Gamma}^{I}\phi_{c}^{I}\Psi_{zd}{f^{cdb}}_{a}
+2i\overline{\epsilon}\Phi_{zc}\tilde{\lambda}_{d}{f^{cdb}}_{a}\\
\label{brstbos6}
\delta\tilde{A}_{\overline{z}a}^{b}
&=
-2i\overline{\tilde{\epsilon}}
\hat{\Gamma}^{I}\phi_{c}^{I}\tilde{\Psi}_{\overline{z}d}{f^{cdb}}_{a}
+2i\overline{\tilde{\epsilon}}\Phi_{\overline{z}c}
\psi_{d}{f^{cdb}}_{a}.
\end{align}

\subsection{Derivation of quantum mechanics}
\label{cy2sec2}
Now we consider the reduction 
to a low-energy effective one-dimensional field theory on $\mathbb{R}$, 
that is membrane quantum mechanics. 
As the size of the Riemann surface shrinks, 
only the light degrees of freedom are relevant. 
To keep track of them  
we have to find the static configurations that minimize the energy, 
that is the zero-energy conditions. 
We can replace the zero-energy conditions by a set of BPS equations. 
In addition, we set all the fermionic fields to zero 
because we are interested in bosonic BPS configurations. 
Then the BPS equations, which correspond to the vanishing conditions of the 
BRST transformations 
(\ref{brstfermi1})-(\ref{brstfermi4}) 
for the fermionic fields, are
\begin{align}
\label{bps1}
&D_{z}\phi^{I}=0,\ \ \ 
D_{\overline{z}}\phi^{I}=0\\
\label{bps2}
&D_{z}\Phi_{\overline{z}}=0,\ \ \ D_{\overline{z}}\Phi_{z}=0\\
\label{bps3}
&[\phi^{I},\phi^{J},\phi^{K}]=0\\
\label{bps4}
&[\Phi_{z},\Phi_{\overline{z}},\phi^{I}]=0,\ \ \ 
[\Phi_{z},\phi^{I},\phi^{J}]=0,\ \ \ 
[\Phi_{\overline{z}},\phi^{I},\phi^{J}]=0\\
\label{bps5}
&[\Phi_{w},\Phi_{\overline{w}},\Phi_{z}]=0,\ \ \ 
[\Phi_{\overline{w}},\Phi_{w},\Phi_{\overline{z}}]=0.
\end{align}

We first note that according to the algebraic equations 
(\ref{bps3}), (\ref{bps4}) and (\ref{bps5}),  
all the bosonic Higgs fields have to lie in the same plane 
in the $SO(4)$ gauge group. 
Thus we can write them as
\begin{align}
\label{bpsk3a}
\phi^{I}=(\phi^{I1},\phi^{I2},0,0)^{T},\ \ \ \ \ 
\Phi_{z}=(\Phi_{z}^{1},\Phi_{z}^{2},0,0)^{T},\ \ \ \ \ 
\Phi_{\overline{z}}
=(\Phi_{\overline{z}}^{1},\Phi_{\overline{z}}^{2},0,0)^{T}.
\end{align} 
Correspondingly via supersymmetry one can also write the fermionic
partners as
\begin{align}
\label{bpsk3b}
\psi&=(\psi^{1},\psi^{2},0,0)^{T},& 
\tilde{\lambda}&=(\tilde{\lambda}^{1},\tilde{\lambda}^{2},0,0)^{T}\\
\label{bpsk3b1}
\Psi_{z}&=(\Psi_{z}^{1},\Psi_{z}^{2},0,0)^{T},&
\tilde{\Psi}_{\overline{z}}&=(\tilde{\Psi}_{\overline{z}}^{1},\tilde{\Psi}_{\overline{z}}^{2},0,0)^{T}.
\end{align} 
The configurations (\ref{bpsk3a})-(\ref{bpsk3b1}) 
generically break the original 
$SO(4)$ gauge group down to $U(1)\times U(1)$. 
Taking into account these solutions and 
the BPS equations (\ref{bps1}), (\ref{bps2}) 
we find that $\tilde{A}_{z3}^{1}=\tilde{A}_{z3}^{2}=
\tilde{A}_{z4}^{1}=\tilde{A}_{z4}^{2}=0$. 
This implies that 
these components of the gauge field now become massive 
by the Higgs mechanism. 
Then we should follow the time evolution for 
remaining degrees of freedom in the low-energy effecvie theory.

To achieve this consistently 
we further need to impose the Gauss law constraint. 
This requires that 
the gauge field is flat; 
$\tilde{F}_{z\overline{z}}=0$. 
Recall that we are now considering the case where 
the genus of the Riemann surface is greater than one. 
In that case the generic flat connections are irreducible.  
As long as we only consider 
irreducible flat connections, the Laplacian has no zero modes. 
Accordingly it is not allowed for scalar fields to have non-trivial
values and it is required that $\phi^{I}=0$ 
\footnote{
Such BPS solutions with the irreducible connections 
have been considered in 
the four-dimensional topologically twisted Yang-Mills theories 
defined on the product of two Riemann surfaces  
\cite{Bershadsky:1995vm,Kapustin:2006pk,Kapustin:2006hi} and  
the corresponding decoupling limit for the brane description 
has been argued in \cite{Maldacena:2000mw}.}. 
To sum up, the above set of equations over the compact Riemann surface 
reduces to
\begin{align}
\label{bps01}
&\tilde{F}_{z\overline{z}2}^{1}=0\\
\label{bps02}
&\partial_{\overline{z}}\Phi_{z1}
+\tilde{A}_{\overline{z}2}^{1}\Phi_{z2}=0\\
\label{bps03}
&\partial_{\overline{z}}\Phi_{z2}
-\tilde{A}_{\overline{z}2}^{1}\Phi_{z1}=0.
\end{align}

Let us discuss the generic BPS configuration obeying 
(\ref{bps01})-(\ref{bps03}). 
Since we are now considering a compact Riemann surface 
of genus $g$,  
there are $g$ holomorphic $(1,0)$-forms $\omega_{i}$, $i=1,\cdots,g$ 
and $g$ anti-holomorphic
$(0,1)$-forms $\overline{\omega}_{i}$. 
Let us normalize them as
\begin{align}
\label{period1}
\int_{a_{i}}\omega_{j}=\delta_{ij},\ \ \ 
\int_{b_{i}}\omega_{j}=\Omega_{ij}
\end{align}
with $a_{i}$, $b_{i}$ being canonical homology basis 
for $H_{1}(\Sigma_{g})$. 
The matrix $\Omega$ is the period matrix 
of the Riemann surface. 
It is a $g\times g$ complex symmetric matrix with positive 
imaginary part. 
The equation (\ref{bps01}) imposes the flatness condition for  
the $U(1)$ gauge field $\tilde{A}_{z2}^{1}$. 
The space of the $U(1)$ flat connection on a compact Riemann surface 
is the torus known as the Jacobi variety denoted by 
$\mathrm{Jac}(\Sigma_{g})$. 
The flat gauge fields can be expressed in the form 
\cite{AlvarezGaume:1986es}
\begin{align}
\label{az1}
\tilde{A}_{z2}^{1}
&=-2\pi \sum_{i,j=1}^{g}
\left(
\Omega-\overline{\Omega}
\right)^{-1}_{ij}
\Theta^{i}\omega_{j}, \ \ \ \ \ 
\tilde{A}_{\overline{z}2}^{1}
=2\pi \sum_{i,j=1}^{g}
\left(
\Omega-\overline{\Omega}
\right)^{-1}_{ij}
\overline{\Theta}^{i}\overline{\omega}_{j}
\end{align}
where $\Theta^{i}:=\zeta^{i}+\overline{\Omega}_{ij}\xi^{j}$ 
represents the complex coordinate of $\mathrm{Jac}(\Sigma_{g})$ which 
characterizes the twists  
$e^{2\pi i\xi^{i}}$ and 
$e^{-2\pi i\zeta^{i}}$ 
around the $i$-th homology cycles $a_{i}$ and $b_{i}$. 
Notice that $\xi^{i}\rightarrow \xi^{i}+m^{i}$, 
$\zeta^{i}\rightarrow \zeta^{i}+n^{i}$ 
for $n^{i}, m^{i}\in \mathbb{Z}$ 
gives rise to the same point on $\mathrm{Jac}(\Sigma_{g})$. 
This implies that 
$\mathrm{Jac}(\Sigma_{g})=\mathbb{C}^{g}/L_{\Omega}$ 
where $L_{\Omega}$ is the lattice generated 
by $\mathbb{Z}^{g}+\Omega\mathbb{Z}^{g}$. 
We define a function
\begin{align}
\label{phifct1}
\varphi:=-2\pi\sum_{i,j=1}^{g}\left(
\Omega-\overline{\Omega}
\right)^{-1}_{ij}
\left(
\Theta^{i}f_{j}(z)
-\overline{\Theta}^{i}\overline{f}_{j}(\overline{z})
\right)
\end{align}
where $f_{i}(z):=\int^{z}\omega_{i}$ is the holomorphic 
function of $z$ that obeys the relations 
$f_{i}|_{a_{j}}=\delta_{ij}$ and $f_{i}|_{b_{j}}=\Omega_{ij}$. 
Then we can write the flat gauge fields as
\begin{align}
\label{az2}
\tilde{A}_{z2}^{1}=\partial_{z}\varphi, \ \ \ \ \ \ \ \ \ \ 
\tilde{A}_{\overline{z}2}^{1}=\partial_{\overline{z}}\varphi. 
\end{align}
Using the above expressions for the $U(1)$ flat connection, 
the generic solutions to the equation (\ref{bps02}) and (\ref{bps03}) 
can be expressed as
\begin{align}
\label{bosconf1b}
\Phi_{z1}(z,\overline{z})-i\Phi_{z2}(z,\overline{z})
=&e^{-i\varphi(z,\overline{z})}\sum_{i=1}^{g}x_{A}^{i}\omega_{i}
\nonumber\\
\Phi_{z1}(z,\overline{z})+i\Phi_{z2}(z,\overline{z})
=&e^{i\varphi(z,\overline{z})}\sum_{i=1}^{g}x_{B}^{i}\omega_{i}
\end{align}
where $x_{A}^{i}$, $x_{B}^{i} \in \mathbb{C}$ 
are constant on the Riemann surface. 
Since we take the limit 
where the Riemann surface $\Sigma_{g}$ shrinks to zero size, 
the space-time configurations of the membranes 
should be expressed as single-valued functions 
of $z$ and $\overline{z}$ 
in the low-energy effective quantum mechanics. 
In other words, 
$\xi^{i}$ and $\zeta^{i}$ can only be integers 
and therefore the $U(1)$ flat gauge fields 
$\tilde{A}_{z2}^{1}$ and $\tilde{A}_{\overline{z}2}^{1}$ are quantized. 
The single-valuedness condition requires 
that the point of the $\mathrm{Jac}(\Sigma_{g})$ 
is fixed.

Putting all together, the general bosonic BPS 
configurations are given by
\begin{align}
\label{bpsconf1}
&\phi^{I}=0\nonumber\\
&\Phi_{z}=\sum_{i=1}^{g}
\left(\begin{array}{c}
\frac12\left(
e^{-i\varphi}x_{A}^{i}
+e^{i\varphi}x_{B}^{i}
\right)\\
\frac{i}{2}
\left(e^{-i\varphi}x_{A}^{i}-e^{i\varphi}x_{B}^{i}
\right)
\\
0\\
0\\
\end{array}
\right)\omega_{i},\ \ \ 
\Phi_{\overline{z}}=
\sum_{i=1}^{g}
\left(\begin{array}{c}
\frac12\left(
e^{i\varphi}\overline{x}_{A}^{i}
+e^{-i\varphi}\overline{x}_{B}^{i}
\right)\\
-\frac{i}{2}
\left(e^{i\varphi}\overline{x}_{A}^{i}
-e^{-i\varphi}\overline{x}_{B}^{i}
\right)
\\
0\\
0\\
\end{array}
\right)
\overline{\omega}_{i}
\nonumber\\
&\tilde{A}_{z}=\left(
\begin{array}{cccc}
0&\partial_{z}\varphi(z,\overline{z})&0&0\\
-\partial_{z}\varphi(z,\overline{z})&0&0&0\\
0&0&0&\tilde{A}_{z4}^{3}(z,\overline{z})\\
0&0&-\tilde{A}_{z4}^{3}(z,\overline{z})&0\\
\end{array}
\right)
\end{align}
where $\tilde{A}_{z4}^{3}$ and 
$\tilde{A}^{3}_{\overline{z}4}$ are the Abelian gauge fields 
associated with preserved $U(1)$ symmetry and they do not receive 
any constraints from the BPS conditions.

By virtue of the supersymmetry 
we can write the corresponding fermionic fields
from the bosonic configurations (\ref{bpsconf1}) as
\begin{align}
\label{bpsconf2}
\psi&=0,&
\tilde{\lambda}&=0\nonumber\\
\Psi_{z}
&=\sum_{i=1}^{g}
\left(
\begin{array}{c}
\frac12\left(
\Psi_{A}^{i}+\Psi_{B}^{i}
\right)\\
\frac{i}{2}\left(
\Psi_{A}^{i}-\Psi_{B}^{i}
\right)\\
0\\
0\\
\end{array}
\right)
\omega_{i}
,& 
\tilde{\Psi}_{\overline{z}}
&=\sum_{i=1}^{g}
\left(
\begin{array}{c}
\frac12 \left(
\tilde{\Psi}_{A}^{i}+\tilde{\Psi}_{B}^{i}
\right)\\
-\frac{i}{2} \left(
\tilde{\Psi}_{A}^{i}-\tilde{\Psi}_{B}^{i}\right)\\
0\\
0\\
\end{array}
\right)\overline{\omega}_{i}.
\end{align}

Substituting the BPS configuration (\ref{bpsconf1}) 
and (\ref{bpsconf2}) 
into the twisted action (\ref{twistedlagrangian}), we find
\begin{align}
\label{twistbpsaction1}
S=\int_{\mathbb{R}}dt\int_{\Sigma_{g}}d^{2}z
\Biggl[
\left(
D_{0}\Phi_{z}^{a},D_{0}\Phi_{\overline{z}a}
\right)
+
\left(
\overline{\Psi}_{z}^{a},D_{0}\tilde{\Psi}_{\overline{z}a}
\right)
-
\left(
\overline{\tilde{\Psi}}_{\overline{z}}^{a},D_{0}\Psi_{za}
\right)
\nonumber\\
-\frac{k}{2\pi}\tilde{A}_{02}^{1}\tilde{F}_{z\overline{z}4}^{3}
-\frac{k}{4\pi}
\left(
\tilde{A}_{z2}^{1}\dot{\tilde{A}}_{\overline{z}4}^{3}
-\tilde{A}_{\overline{z}2}^{1}\dot{\tilde{A}}_{z4}^{3}
\right)
\Biggr].
\end{align}
Since the gauge fields $\tilde{A}_{z2}^{1}$, $\tilde{A}_{\overline{z}2}^{1}$ 
are quantized and their time derivatives 
do not show up in the effective action, 
they can be integrated out as the auxiliary fields. 
They give rise to the constraints 
$\dot{\tilde{A}}_{z4}^{3}=\dot{\tilde{A}}_{\overline{z}4}^{3}=0$. 

Making use of the Riemann bilinear relation \cite{MR1139765} 
\begin{align}
\label{bil01}
\int_{\Sigma_{g}}\omega \wedge \eta
=\sum_{i=1}^{g}
\left[
\int_{a_{i}}\omega\int_{b_{i}}\eta
-\int_{b_{i}}\omega\int_{a_{i}}\eta
\right]
\end{align}
and performing the integration on $\Sigma_{g}$ 
we obtain the low-energy effective gauged quantum mechanics
\begin{align}
\label{eff001}
S=\int_{\mathbb{R}}dt 
\Biggl[
&
\sum_{i,j}
\left(
\mathrm{Im}\ \Omega
\right)_{ij}
\left(
D_{0}x^{ia}D_{0}\overline{x}_{a}^{j}
+\overline{\Psi}^{ia}D_{0}\tilde{\Psi}_{a}^{j}
-\overline{\tilde{\Psi}}^{ia}D_{0}\Psi_{a}^{j}
\right)
-kC_{1}(E)\tilde{A}_{02}^{1}
\Biggr].
\end{align}
Here the indices $a=A,B$ stand for the 
two internal degrees of freedom for the two M2-branes. 
The covariant derivatives are defined by
\begin{align}
D_{0}x_{A}^{i}&=\dot{x}_{A}^{i}+i\tilde{A}^{1}_{02}x_{A}^{i}, 
&D_{0}x_{B}^{i}&=\dot{x}_{B}^{i}-i\tilde{A}^{1}_{02}x_{B}^{i}\\
D_{0}\Psi_{A}^{i}&=\dot{\Psi}_{A}^{i}+i\tilde{A}_{02}^{1}\Psi_{A}^{i}, 
&D_{0}\Psi_{B}^{i}&=\dot{\Psi}_{B}^{i}-i\tilde{A}_{02}^{1}\Psi_{B}^{i}\\
D_{0}\tilde{\Psi}_{A}^{i}&=\dot{\tilde{\Psi}}_{A}^{i}
-i\tilde{A}_{02}^{1}\tilde{\Psi}_{A}^{i}, 
&D_{0}\Psi_{B}^{i}&=\dot{\tilde{\Psi}}_{B}^{i}
+i\tilde{A}_{02}^{1}\tilde{\Psi}_{B}^{i}
\end{align}
and the Chern number $C_{1}(E) \in \mathbb{Z}$ 
is associated to the $U(1)$ principal bundle $E\rightarrow \Sigma_{g}$
over the Riemann surface
\begin{align}
\label{ch03}
C_{1}(E)=\int_{\Sigma_{g}}c_{1}(E)
=\frac{1}{2\pi}\int_{\Sigma_{g}}d^{2}z \tilde{F}_{z\overline{z}4}^{3}.
\end{align}
The action (\ref{eff001}) has the invariance 
under the one-dimensional $SL(2,\mathbb{R})$ conformal transformations
\begin{align}
\label{k3qmconf1}
\delta t&=f(t)=a+bt+ct^{2}, 
&\delta\partial_{0}&=-\dot{f}\partial_{0}\\
\delta x_{a}^{i}&=\frac12 \dot{f} x_{a}^{i}, 
&\delta \tilde{A}_{02}^{1}&=-\dot{f}\tilde{A}_{02}^{1}\\
\delta \Psi^{i}_{a}&=0, 
&\delta \tilde{\Psi}^{i}_{a}&=0.
\end{align}
Also the action (\ref{eff001}) is invariant 
under the $\mathcal{N}=8$ supersymmetry transformations
\begin{align}
\label{k3qmsusy1}
\delta x^{i}_{a}&=2i\overline{\tilde{\epsilon}}\Psi_{a}^{i}, 
&\delta
 \overline{x}^{i}_{a}&=2i\overline{\epsilon}\tilde{\Psi}_{a}^{i}\\
\delta\Psi_{a}^{i}&=-iD_{0}x_{a}^{i}\epsilon, 
&\delta\tilde{\Psi}_{a}^{i}&=iD_{0}\overline{x}_{a}^{i}\tilde{\epsilon}\\
\delta\tilde{A}_{02}^{1}&=0.
\end{align}

Therefore we conclude that 
the $\mathcal{N}=8$ superconformal gauged quantum mechanics 
(\ref{eff001}) may describe the low-energy effective motion 
of the two wrapped M2-branes around $\Sigma_{g}$ 
probing a K3 surface.

As seen from the action (\ref{eff001}), 
the $U(1)$ gauge field $\tilde{A}_{02}^{1}$,  
due to the absence of the kinetic term, 
is regarded as an auxiliary field. 
In consequence the gauge field has no contribution to the Hamiltonian. 
Hence the corresponding gauge symmetry yields an integral of motion as a
moment map $\mu:\mathcal{M}\rightarrow \mathfrak{u}(1)^{*}$ 
and we can reduce the phase space $\mathcal{M}$ 
to $\mathcal{M}_{c}=\mu^{-1}(c)$ by fixing the inverse of 
the moment map at a point $c\in \mathfrak{u}(1)^{*}$. 
Choosing a temporal gauge $\tilde{A}_{02}^{1}=0$, 
we find the action
\begin{align}
\label{eff0001}
S=\int_{\mathbb{R}} dt 
\sum_{i,j}\left(
\mathrm{Im}\Omega
\right)_{ij}\left(
\dot{x}^{ia}\dot{\overline{x}}^{j}_{a}
+\overline{\Psi}^{ia}\dot{\tilde{\Psi}}^{j}_{a}
-\overline{\tilde{\Psi}}^{ia}\dot{\Psi}^{j}_{a}
\right)
\end{align}
and the Gauss law constraint
\begin{align}
\label{k3gauss1}
\phi_{0}:=
kC_{1}(E)
+i\sum_{i,j}\left(\textrm{Im}\Omega\right)_{ij}
\left[
K_{ij}
+2\left(
\overline{\Psi}^{i}_{A}\tilde{\Psi}_{A}^{j}
-\overline{\Psi}_{B}^{i}\tilde{\Psi}_{B}^{j}
\right)
\right]
=0
\end{align}
where 
\begin{align}
\label{kconst}
K_{ij}:=
\left(
\dot{x}^{i}_{A}\overline{x}^{j}_{A}
-x_{A}^{i}\dot{\overline{x}}_{A}^{j}
\right)
-
\left(
\dot{x}^{i}_{B}\overline{x}^{j}_{B}
-x^{i}_{B}\dot{\overline{x}}_{B}^{j}
\right).
\end{align} 
The constraint equation (\ref{k3gauss1}) 
requires that all states in the Hilbert space are gauge invariant. 
In this case the symmetry of 
the system is not so large as in 
the previous superconformal gauged quantum mechanical models 
(\ref{effs1}) and (\ref{effss2}). 
It is curious to know whether 
the superconformal gauged quantum mechanics 
(\ref{eff001}) (or (\ref{eff0001}) together with (\ref{k3gauss1})) 
have a reduced Lagrangian description with 
an inverse-square type potential. 
However, our result may 
drop a hint on the obstructed construction of SCQM that 
a large class of SCQM could be formulated as 
``gauged quantum mechanics'' with the help of auxiliary gauge fields 
as in \cite{MR0478225,Polychronakos:1991bx,Fedoruk:2008hk}.

\section{Conclusion and discussion}
\label{secconc}
We have studied the IR quantum mechanics 
resulting from the multiple M2-branes 
wrapping a compact Riemann surface $\Sigma_{g}$ 
after shrinking the size of the Riemann surface 
by reducing the BLG-model and the ABJM-model. 
For $g=1$ the dimensional reductions of the BLG-model 
and the ABJM-model yield the low-energy effective 
$\mathcal{N}=16$ and $\mathcal{N}=12$ 
superconformal gauged quantum mechanical models respectively. 
After the integration of the auxiliary gauge fields, 
$OSp(16|2)$ quantum mechanics (\ref{effk1}) 
and $SU(1,1|6)$ quantum mechanics (\ref{abjmscqm0}) emerge 
from the reduced theories. 
%
For $g\neq 1$ the Riemann surface is singled out as 
a calibrated holomorphic curve in a Calabi-Yau manifold  
to preserve supersymmetry. 
The IR quantum mechanical models have 
$\mathcal{N}=8$, $4$, $2$ and $2$ supersymmetries for 
K3, $CY_{3}$, $CY_{4}$ and $CY_{5}$ respectively. 
When the Calabi-Yau manifolds are 
constructed via decomposable line bundles 
over the Riemann surface,  
the K3 surface essentially allows for a unique topological twist  
while for the other Calabi-Yau manifolds 
there are infinitely many topological twists 
which are specified by the degrees of the line bundles. 
In particular we have analyzed the 
two wrapped M2-branes around a holomorphic genus $g>1$ curve 
exploring a K3 surface based on the topologically twisted BLG-model. 
We have found 
the $\mathcal{N}=8$ superconformal gauged quantum mechanics
(\ref{eff001}) that 
may describe the low-energy dynamics of the wrapped M2-branes 
in a K3 surface. 
It is known that \cite{MR0478225,Polychronakos:1991bx,Fedoruk:2008hk} 
there are the connections of the gauged quantum mechanics 
to the conformal mechanical models, the Calogero model and their
generalizations. 
An interesting question is what type of interaction potential, 
if it exists, 
may characterize our superconformal gauged quantum mehcanics 
(\ref{eff001}). 
This remains open issue for future investigation.  

There are a number of future aspects of the present work. 
In particular, they contain the following impressive subjects:

\begin{enumerate}
\item AdS$_{2}/$CFT$_{1}$ correspondence

One of the most appealing programs relevant to our work 
is to attack the $\textrm{AdS}_{2}/\textrm{CFT}_{1}$ correspondence. 
This is the most significant case of 
$\textrm{AdS}_{d+1}/\textrm{CFT}_{d}$ correspondence 
\cite{Maldacena:1997re}
in that all known extremal black holes contain 
the AdS$_{2}$ factor in their near horizon geometries.

It has been discussed 
in \cite{Claus:1998ts,Gibbons:1998fa} that 
the motion of the particle 
near the horizon of the extreme Reissner-Nordstr\"{o}m 
black hole is described by the (super)conformal mechanics. 
Since such black holes can be alternatively described 
by the wrapped M2-branes around a compact Riemann surface in M-theory, 
we expect that our superconformal quantum mechanics 
provides further examples and the M-theoretic interpretation.

It has been pointed out in \cite{Chamon:2011xk,Jackiw:2012ur} 
that the correlation functions of 
the DFF-model \cite{deAlfaro:1976je} 
have the expected scaling behaviors 
although one cannot assume the existence of the normalized and conformal
invariant vacuum states in conformal quantum mechanics 
as in other higher dimensional conformal field theories. 
We would like to extend the analysis 
to superconformal quantum mechanics including our models. 

\item Indices and the reduced Gromov-Witten invariants

The formula for the numbers of genus $g$ curves in a K3 surface, 
the so-called reduced Gromov-Witten invariants \cite{MR1750955} 
has been firstly proposed by Yau and Zaslow in the analysis of 
the wrapped D3-brane \cite{Yau:1995mv}. 
Closely related to their setup, 
our $\mathcal{N}=8$ superconformal gauged quantum mechanics 
(\ref{eff001}) appears from the wrapped M2-branes instead of the D3-brane. 
It would be interesting to compute the indices and 
to extract enumerative information and structure from our model.

\item 1d-2d relation

In analogy with the fascinating stories 
arising from the compactification of M5-branes, 
for example, the AGT-relation \cite{Alday:2009aq}, 
the DGG-relation \cite{Dimofte:2011ju} 
and the 2d-4d relation \cite{Gadde:2013sca}, 
it would be attractive 
to find the relationship between 
the superconformal field theories and 
the geometries or relevant dualities from M2-branes, 
i.e. ``1d-2d relation''. 
It has been observed in \cite{Bellucci:2004wn} that 
the WDVV equation \cite{Witten:1989ig,Dijkgraaf:1990dj} 
and the twisted periods \cite{MR2070050,MR2999308} 
which are relevant to two-dimensional geometries 
and topological field theories appear 
from the constraint conditions 
for the constructions of $\mathcal{N}=4$ superconformal mechanics. 
It would be interesting to investigate 
whether our M-theoretical construction of 
superconformal quantum mechanics 
could help to understand and generalize such relations.

\end{enumerate}

\subsection*{Acknowledgments}
I am deeply indebted to Hirosi Ooguri 
for numerous discussions, 
valuable comments and stimulating suggestions 
on this project. 
I am grateful to Yu Nakayama and Satoshi Yamaguchi 
for helpful discussions and 
for sharing their insights throughout the course of this work.  
I would also like to thank 
Abhijit Gadde, Kazuo Hosomichi, Daniel L. Jafferis, 
Kazunobu Maruyoshi, Takuya Okuda, Du Pei, 
Pavel Putrov and Yuji Tachikawa for useful discussions and comments. 
This work was supported 
in part by JSPS fellowships for Young Scientists.


\bibliographystyle{utphys}
\bibliography{ref}

\end{document}